\documentclass[twocolumn, graphics, floatfix, a4paper, aps, prb,
longbibliography, superscriptaddress, noeprint]{revtex4-2}

\usepackage{graphicx}
\usepackage{dcolumn}
\usepackage{bm}

\usepackage{amsmath}
\usepackage{amssymb}
\usepackage{float} 
\usepackage{graphicx}
\usepackage{blkarray}
\usepackage{nicematrix}
\usepackage[colorinlistoftodos]{todonotes}
\usepackage[colorlinks=true, allcolors=blue]{hyperref}
\usepackage{multirow}
\usepackage{gensymb}
\usepackage{braket}
\usepackage{bm}
\usepackage{subcaption}
\usepackage{outlines}
\usepackage{leftindex}
\usepackage[normalem]{ulem}
\usepackage[newcommands]{ragged2e}
\usepackage[capitalize]{cleveref}

\usepackage{listings}
\usepackage{color} 
\definecolor{mygreen}{RGB}{28,172,0} 
\definecolor{mylilas}{RGB}{170,55,241}

\usepackage{xcolor}

\definecolor{codegreen}{rgb}{0,0.6,0}
\definecolor{codegray}{rgb}{0.5,0.5,0.5}
\definecolor{codepurple}{rgb}{0.58,0,0.82}
\definecolor{backcolour}{rgb}{0.95,0.95,0.92}

\lstdefinestyle{codestyle}{
	backgroundcolor=\color{backcolour},   
	commentstyle=\color{codegreen},
	keywordstyle=\color{magenta},
	numberstyle=\tiny\color{codegray},
	stringstyle=\color{codepurple},
	basicstyle=\ttfamily\footnotesize,
	breakatwhitespace=false,         
	breaklines=true,                 
	captionpos=b,                    
	keepspaces=true,                 
	numbers=left,                    
	numbersep=5pt,                  
	showspaces=false,                
	showstringspaces=false,
	showtabs=false,                  
	tabsize=2
}

\renewcommand{\Re}{\text{Re}}
\renewcommand{\Im}{\text{Im}}

\newcommand{\bk}{\bm{k}}

\newcommand{\intd}[1]{\text{ d} #1}

\newcommand{\exval}[1]{\langle #1 \rangle}

\definecolor{EL-color}{named}{blue}
\definecolor{PT-color}{named}{orange}
\definecolor{PT-colorC}{named}{magenta}
\definecolor{SP-color}{named}{purple}

\newcommand{\co}[3]{\hat{c}^{\vphantom{\dagger}}_{#1 #2 #3}}
\newcommand{\cod}[3]{\hat{c}^\dagger_{#1 #2 #3}}

\newcommand{\cspinor}[1]{\hat{\bm{c}}^{\vphantom{\dagger}}_{#1}}
\newcommand{\cdspinor}[1]{\hat{\bm{c}}^\dagger_{#1}}

\newcommand{\bo}[4]{\hat{b}^{\vphantom{\dagger}}_{#1 #2 #3 #4}}
\newcommand{\bod}[4]{\hat{b}^\dagger_{#1 #2 #3 #4}}

\newcommand{\ho}[4]{\hat{h}^{\vphantom{\dagger}}_{#1 #2 #3 #4}}
\newcommand{\hod}[4]{\hat{h}^\dagger_{#1 #2 #3 #4}}

\newcommand{\deltareal}{\Delta_{i \alpha j \beta}}

\newcommand{\qpdw}{\bm{q}}
\newcommand{\omegahessian}{\partial^2_{\Delta} \Omega}

\newcount\colveccount
\newcommand*\colvec[1]{
	\global\colveccount#1
	\begin{pmatrix}
		\colvecnext
	}
	\def\colvecnext#1{
		#1
		\global\advance\colveccount-1
		\ifnum\colveccount>0
		\\
		\expandafter\colvecnext
		\else
	\end{pmatrix}
	\fi
}
\newcommand{\bpm}{\begin{pmatrix}}
\newcommand{\epm}{\end{pmatrix}}

\newcommand{\bea}{\begin{equation} \begin{aligned}}
\newcommand{\eea}{\end{aligned} \end{equation} }

\begin{document}



\title{Superconductivity and pair density waves from nearest-neighbor interactions in frustrated lattice geometries}

\author{E.O.~Lamponen}
\affiliation{Department of Applied Physics, Aalto University School of Science, FI-00076 Aalto, Finland}
\author{S.K.~Pöntys}%
\affiliation{Department of Applied Physics, Aalto University School of Science, FI-00076 Aalto, Finland}
\author{P.~Törmä}
\thanks{paivi.torma@aalto.fi}
\affiliation{Department of Applied Physics, Aalto University School of Science, FI-00076 Aalto, Finland}





\begin{abstract}
We consider superconductivity and pair density waves (PDWs) arising from off-site pairing in frustrated lattice geometries. We express the pair susceptibility in a generic form that highlights the importance of both the density of states, and the quantum geometry of the eigenstates and calculate the superfluid weight (stiffness) as well as the Berezinskii-Kosterlitz-Thouless (BKT) temperature. Paradigmatic bipartite (Lieb) and non-bipartite (kagome) lattices are studied as examples. For bipartite lattices, nearest-neighbor pairing vanishes in a flat band. In the Lieb lattice flat band, we find a PDW at a finite interaction and show that its pair wave vector is determined by the quantum geometry of the band. In the kagome flat band, nearest-neighbor pairing is possible for infinitesimal interactions. At the kagome van Hove singularity, the pair susceptibility predicts a PDW due to sublattice interference, however, we find that its stiffness is zero due to the shape of the Fermi surface. Our results indicate that nearest-neighbor pairing at flat band and van Hove singularities is strongly influenced by the geometric properties of the eigenfunctions, and it is crucial to determine the superfluid weight of the superconducting and PDW orders as it may contradict the predictions by pairing susceptibility. 
\end{abstract}

\maketitle



\section{Introduction}

Pair formation and superconductivity are known to be strongly affected by the density of states (DOS) of the electrons. Recently, the geometric structure of the Bloch states has been identified as another key factor determining properties of quantum materials~\cite{yu2024quantumgeometryquantummaterials}. For example, the quantum geometry of the Bloch states allows superconductivity in a flat band~\cite{peottaSuperfluidityTopologicallyNontrivial2015, julkuGeometricOriginSuperfluidity2016, liangBandGeometryBerry2017, Torma2022ReviewQuantumGeometry, Huhtinen2022FlatBandSCQuantumMetric} where the high DOS favors pairing but single particles seem localized and one would naively expect no supercurrent. This prediction may be relevant for understanding superconductivity in twisted bilayer graphene~\cite{julkuSuperfluidWeightBerezinskiiKosterlitzThouless2020, Hu2019MATBGSW,Xie2020TopologyBoundSCTBG,Torma2022ReviewQuantumGeometry,Tian2023QuantumGeoSC,tanakaSuperfluidStiffnessMagicangle2025} and other flat band systems~\cite{regnault_Catalogue_2022,jiang20242dtheoreticallytwistablematerial}. 
Also PDW~\cite{agterbergPhysicsPairDensity2020, settyMechanismFluctuatingPair2023, wu_pairdw_2023, loder_sc_2010, yerinMultipleqCurrentStates2023a} order has been analyzed in terms of quantum geometry~\cite{fuExoticChargeDensity2025, chenPairDensityWave2023, jiangPairDensityWaves2023, ticeaPairDensityWave2024, wangDensityMatrixRenormalization2025, kitamuraQuantumGeometricEffect2022}. The vast majority of theory research on flat band superconductivity considers local (on-site) pairing~\cite{heikkilaFlatBandsTopological2011, kopninHightemperatureSurfaceSuperconductivity2011, peottaSuperfluidityTopologicallyNontrivial2015, julkuGeometricOriginSuperfluidity2016, liangBandGeometryBerry2017, Huhtinen2022FlatBandSCQuantumMetric,Torma2022ReviewQuantumGeometry,Peotta2023review,Iskin_2021_TwoBodyMultiLatQMgeom,BatrouniChan2021,Chan2022BandTouching}, with a few exceptions~\cite{julkuSuperfluidWeightBerezinskiiKosterlitzThouless2020,Powell2022tRVBFlatBand,Braz2024LongRangeIntFlatBand,BlackSchaffer2022NN_NNNinteractionFlatBand,Wang2024LongRangeIntsFlatBand,Hofmann2023SC_CWD,Peri2021TBGFragileAndSC,Hofmann2020_SC}. 
Here, we study nearest-neighbor pairing in frustrated lattice systems and highlight the importance of considering the DOS and the eigenfunctions' structure, i.e., quantum geometry, on equal footing.

Dirac points, van Hove singularities (vHs), and flat bands exemplify electronic states with vanishing or diverging DOS. At Dirac points the DOS is zero and consequently one observes quantum critical pairing where superconductivity emerges only beyond a critical attractive interaction strength \cite{sorellaSemiMetalInsulatorTransitionHubbard1992, uchoaSuperconductingStatesPure2007, black-schafferResonatingValenceBonds2007, leeAttractiveHubbardModel2009, limStronglyInteractingTwodimensional2009, mazzucchiSemimetalSuperfluidQuantum2013, parisentoldinFermionicQuantumCriticality2015}. 
Previous studies of Dirac points and vHs have mainly focused on DOS effects that suppress or enhance pairing, 
although the structure of the eigenstates may affect pair formation via so-called sublattice interference \cite{kieselSublatticeInterferenceKagome2012, wuSublatticeInterferencePromotes2023} (also known as sublattice polarization \cite{linSublatticePolarizationDestructive2024}). In flat-band superconductivity, the impact of quantum geometry on pair transport is well understood, however, its effect on the pairing type and symmetry has not been studied as much~\cite{Daido2024, jahinEnhancedKohnLuttingerTopological2025, sunFlatbandFFLOState2024}. We express the pairing (particle-particle channel) susceptibility in a form that allows one to directly estimate the separate effects of the DOS and the eigenfunctions on the formation of pairs, for a generic interaction and lattice geometry. We point out that, for bipartite lattices, nearest-neighbor pairing (sometimes referred to as resonating valence bond (RVB) -type~\cite{andersonResonatingValenceBond1987, black-schafferResonatingValenceBonds2007, baskaranResonatingValenceBond2017,julkuSuperfluidWeightBerezinskiiKosterlitzThouless2020} pairing) identically vanishes. By numerical studies of the mean-field order parameters and superfluid weight in the Lieb and kagome lattices, we show that attractive nearest-neighbor interactions can lead to quantum critical pairing with superconducting or PDW orders at the vHs and flat bands. The results are compared with on-site pairing schemes. In the Lieb lattice flat band, we find a PDW and show that its character is influenced by the gradients of the orbital compositions of the flat band states, in other words, quantum geometry. In the kagome vHs, PDW order has been previously predicted~\cite{wuSublatticeInterferencePromotes2023,fuExoticChargeDensity2025} with different interaction schemes, however, we find a vanishing superfluid weight for the PDW order in our model. In contrast, the Lieb lattice vHs does not suffer from such effects and shows stable nearest-neighbor pairing comparable to the on-site one.



\section{Pairing susceptibility in \protect\\ mean-field models}

\subsection{Mean-field Hubbard model}

We start with a general multiband Hubbard model Hamiltonian $\hat{H}$ given by
\begin{align}
    \hat{H} &= \sum_{i \alpha, j\beta} \sum_\sigma t_{i \alpha j \beta}^\sigma \cod{i}{\alpha}{\sigma} \co{j}{\beta}{\sigma} - \mu \sum_{i \alpha \sigma} \cod{i}{\alpha}{\sigma} \co{i}{\alpha}{\sigma}
    + \hat{H}_{\text{int}},
    \label{eq: hamiltonian}
\end{align}
where $\cod{i}{\alpha}{\sigma}$ ($\co{i}{\alpha}{\sigma}$) creates (annihilates) a fermion with spin $\sigma$ on orbital $\alpha$ in unit cell $i$. The hopping amplitudes between sites are given by $t_{i \alpha j \beta}^\sigma$, while $\mu$ is the chemical potential. We consider quartic interaction terms of the form 
\begin{align}
    \hat{H}_{\text{int}} = -\sum_{i \alpha, j \beta} J_{i \alpha j \beta} \bod{i}{\alpha}{j}{\beta} \bo{i}{\alpha}{j}{\beta},
    \label{eq: H_int}
\end{align}
where $J_{i \alpha j \beta} > 0$ is the strength of the attractive interaction, $\bo{i}{\alpha}{j}{\beta} = \sum_\sigma A_\sigma \co{i}{\alpha}{\sigma} \co{j}{\beta}{-\sigma}$ is a combination of two annihilation operators with opposite spins, and $A_\sigma$ are real coefficients.  
The operator $\bo{i}{\alpha}{j}{\beta}$ unites several possible interactions into a single form; for example, the standard, local Hubbard-$U$ interaction (with $J_{i \alpha j \beta} = U \delta_{i \alpha, j \beta}, A_\uparrow = 0, A_\downarrow = 1$), as well as long-range density-density ($A_\uparrow = 0, A_\downarrow = 1$) and spin-exhange ($A_\uparrow = 1/2, A_\downarrow = -1/2$) interactions. We use a mean-field approach: $- J_{i \alpha j \beta} \bod{i}{\alpha}{j}{\beta} \bo{i}{\alpha}{j}{\beta} \approx \deltareal^* \bo{i}{\alpha}{j}{\beta} + \text{H.c} + |\deltareal|^2 / J_{i \alpha j \beta}$, where the $\deltareal = -J_{i \alpha j \beta} \exval{\bo{i}{\alpha}{j}{\beta}}$ are the order parameters of the system. We focus solely on the particle-particle channel, i.e. superconducting orders, while ignoring the particle-hole channel (spin and charge orders), even though competing orders could be present \cite{Hofmann2023SC_CWD, scholleSpiralStripeTransition2024, Classen2024, fuExoticChargeDensity2025, wangCompetingElectronicOrders2013}. While care should be taken when interpreting mean-field results, results from more sophisticated methods such as dynamical mean-field theory \cite{penttilaFlatbandRatioQuantum2025}, exact density matrix renormalization group \cite{Chan2022BandTouching} and exact diagonalization \cite{julkuGeometricOriginSuperfluidity2016} have previously been in qualitative agreement with those obtained from mean-field for a variety of lattices, including those with non-isolated flat bands. Later on, we will also go beyond mean field by considering the superfluid weight, i.e. the quadratic stiffness against phase fluctuations of the order parameters. We continue by assuming $J_{i \alpha (i+j) \beta} = J_{0 \alpha j \beta}$ and that the order parameters can only break this translational invariance up to a periodic phase modulation, that is,
\begin{align}
    \Delta_{i \alpha (i+j) \beta} = \overline{\Delta}_{0 \alpha j \beta} e^{i 2 \qpdw \cdot (\bm{r}_{i \alpha} + \bm{r}_{(i+j) \beta}) / 2},
    \label{eq: delta_real_formulation}
\end{align}
where $\overline{\Delta}_{0 \alpha j \beta}$ does not depend on $\qpdw$.
A non-zero $\qpdw$ corresponds to a PDW state. We denote $\bm{r}_{i \alpha} \equiv \bm{R}_{i} + \bm{\delta}_\alpha$, where $\bm{R}_i$ is the position of the $i$'th unit cell, and $\bm{\delta}_\alpha$ is the position of orbital $\alpha$ within the unit cell. 
For this pairing scheme, we derive a Bogoliubov - de Gennes (BdG) Hamiltonian $H_{\text{BdG}}$ (details in App. \ref{sec: supp_bdg_hamiltonian}) such that 

\begin{align}
   \hat{H} &= \sum_{\bm{k}} \cdspinor{\bm{k}} H_{\text{BdG}}(\bm{k}) \cspinor{\bm{k}} + N_c \sum_{j \alpha \beta} \frac{|\overline{\Delta}_{0 \alpha j \beta}|^2}{J_{0 \alpha j \beta}}, \label{eq: H_bdg}\\
   H_{\text{BdG}}(\bm{k}) &= \begin{pmatrix}
       H^\uparrow_{\bm{k} + \qpdw} - \mu \bm{1} & \Delta_{\bm{k}}\\
       \Delta^\dagger_{\bm{k}} & - (H^\downarrow_{-\bm{k} + \qpdw})^* + \mu \bm{1}
   \end{pmatrix}, \label{eq_supp: H_bdg}
\end{align}
where the identity matrix is denoted by $\bm{1}$, the $\bm{k}$-sum is taken over the Brillouin zone, and $N_c$ is the number of unit cells. The pairing matrix $\Delta_{\bm{k}}$ is given by

\begin{align}
    [\Delta_{\bm{k}}]_{\alpha \beta} &= \sum_j \bigg( A_\downarrow \overline{\Delta}_{0 \beta j \alpha} e^{-i \bm{k} \cdot \bm{r}^\Delta_{0 \beta j \alpha}} - A_\uparrow \overline{\Delta}_{0 \alpha j \beta} e^{i \bm{k} \cdot \bm{r}^\Delta_{0 \alpha j \beta}} \bigg).
   \label{eq: delta_k}
\end{align}
Furthermore, $\cspinor{\bm{k}} = (\co{\bm{k} + \qpdw}{\alpha=1}{\uparrow}, \dots, \co{\bm{k} + \qpdw}{\alpha=n_{\text{orb}}}{\uparrow}, \newline \cod{-\bm{k} + \qpdw}{\alpha=1}{\downarrow}, \dots, \cod{-\bm{k} + \qpdw}{\alpha=n_{\text{orb}}}{\downarrow})^T$, while $[H_{\bm{k}}^\sigma]_{\alpha \beta} = \sum_j t_{0 \alpha j \beta}^\sigma e^{i \bm{k} \cdot \bm{r}^\Delta_{0 \alpha j \beta}}$ is the Fourier-transformed kinetic Hamiltonian, with $\co{i}{\alpha}{\sigma} = (1/\sqrt{N_c}) \sum_{\bm{k}} e^{i \bm{k} \cdot \bm{r}_{i \alpha}} \co{\bm{k}}{\alpha}{\sigma}$, $\bm{r}^\Delta_{i \alpha j \beta} \equiv \bm{r}_{j \beta} - \bm{r}_{i \alpha}$ and $n_{\text{orb}}$ the number of orbitals.

We note that the pairing matrix $\Delta_{\bm{k}}$ in Eq. \eqref{eq: delta_k} is in general (i.e. for non-on-site interactions) non-diagonal and $\bm{k}$-dependent. Its structure is of particular interest in this work, as will be seen in the next section.
 
\subsection{Pairing susceptibility}

The equilibrium state of the system is given by the values of $\overline{\Delta}_{0 \alpha j \beta}$ and $\qpdw$ that minimize the grand potential

\begin{align}
   \Omega &= - k_B T \sum_{\bm{k} a} \ln (1 + e^{-E_{\bm{k}a}/ k_B T} ) + N_c \sum_{j \alpha \beta} \frac{|\overline{\Delta}_{0 \alpha j \beta}|^2}{J_{0 \alpha j \beta}},
   \label{eq: omega}
\end{align}
where $E_{\bm{k}a}$ are the $2 n_{\text{orb}}$ eigenvalues of the BdG Hamiltonian $H_{\text{BdG}}(\bm{k})$, $T$ is the temperature, and $k_B$ the Bolzmann constant. We treat the different real-space order parameters as independent variables, thus capturing all possible pairing symmetries (e.g. $s$-wave, $d$-wave etc.) simultaneously. For a given $\qpdw$, the transition from a normal to a superconducting state can be identified by the Hessian matrix $\omegahessian \equiv \text{Hess}(\Omega) |_{\Vec{\Delta} = 0}$,
evaluated at $\Vec{\Delta} = 0$, where $\Vec{\Delta}$ denotes a vector of all independent order parameters. For brevity, we denote an element of $\Vec{\Delta}$ by $\Delta_\mu$, where $\mu\equiv\{\alpha j \beta \}$. The transition from the normal state to a superconducting one occurs when the minimum of $\Omega$ shifts from $\Vec{\Delta} = 0$ to $\Vec{\Delta} \neq 0$ and the smallest eigenvalue of $\omegahessian$ is zero.

We assume a constant interaction strength $J_{0 \alpha j \beta} = J$ for all pairs $0 \alpha, j \beta$ for which it is non-zero; the Hessian becomes $\omegahessian = \frac{2 s N_c}{J} (-\chi + \bm{1})$. Here 
$s = 1$ for on-site interactions, while $s=2$ for off-site interactions when $\Delta_{i\alpha j \beta} = \Delta_{j \beta i \alpha}$ holds; for the general case see App. \ref{sec: supp_pairing_susceptibility}. Here, $\chi$ is the \textit{pairing susceptibility}
\begin{align}
    \chi = -\frac{J}{2 s N_c} \begin{pmatrix}
        \partial_{\Delta^R} \partial_{\Delta^R} \mathcal{E} & 
        \partial_{\Delta^R} \partial_{\Delta^I} \mathcal{E} \\
        \partial_{\Delta^I} \partial_{\Delta^R} \mathcal{E}
        & \partial_{\Delta^I} \partial_{\Delta^I} \mathcal{E}
    \end{pmatrix} ,
\end{align}
with $\mathcal{E} = - k_B T \sum_{\bm{k} a} \ln (1 + e^{-E_{\bm{k}a}/k_B T})$, and $R$ and $I$ denoting the real and imaginary parts of the order parameters, respectively, e.g. $[\partial_{\Delta^R} \partial_{\Delta^I} \mathcal{E}]_{\mu \nu} = \partial_{\Delta^R_\mu} \partial_{\Delta^I_\nu} \mathcal{E}$. The condition for the superconducting transition is now equivalent to the largest eigenvalue of $\chi$ being one. 
The temperature (or interaction) where this happens gives the critical temperature $T_c$ (or critical interaction $J_c$). Utilizing the Hellman-Feynman theorem and assuming time-reversal symmetry (TRS) (App. \ref{sec: supp_pairing_susceptibility}), we obtain the following informative expression for the pairing susceptibility:
\begin{align}
    \chi &= -\frac{J}{2 s N_c} \begin{pmatrix}
                \Re X & \Im X\\
                -\Im X & \Re X
        \end{pmatrix}, \label{eq: full_pair_susceptibility}\\
    X_{\mu \nu} &= \sum_{\bm{k}mn} \frac{n_F(\xi_{\bm{k} + \qpdw m}) + n_F(\xi_{\bm{k} - \qpdw n}) - 1}{\xi_{\bm{k} + \qpdw m} + \xi_{\bm{k} - \qpdw n}} \label{eq: pair_susceptibility} \\
    &\times 2 \braket{m_{\bm{k}+\bm{q}}|\delta \Delta_\mu(\bm{k})|n_{\bm{k}-\bm{q}}} \braket{n_{\bm{k}-\bm{q}}| \delta \Delta_\nu^\dagger(\bm{k}) |m_{\bm{k}+\bm{q}}}.\nonumber
\end{align}
The first line of Eq.~\eqref{eq: pair_susceptibility} depends on the single-particle energies $\epsilon_{\bm{k} m}$ via $\xi_{\bm{k} m} \equiv \epsilon_{\bm{k} m} - \mu$ with $m$ denoting the band, as well as the Fermi-Dirac distribution $n_F$, i.e.~band structure effects such as DOS. The form factor on the second line depends on the corresponding Bloch functions $\ket{m_{\bm{k}}}$ as well as the structure of the interaction via $\delta \Delta_\mu (\bm{k}) \equiv \partial_{\Delta_\mu^R} \Delta_{\bm{k}}$, where $\Delta_{\bm{k}}$ is the pairing matrix given by Eq. \eqref{eq: delta_k}.

At low temperatures, the first line of Eq.~\eqref{eq: pair_susceptibility} becomes increasingly concentrated at the Fermi level, with a divergence in the limit $\xi_{\bm{k} - \qpdw n}, \xi_{\bm{k} + \qpdw m} \rightarrow 0, T \rightarrow 0$. It often dominates the pairing susceptibility, making the band structure near the Fermi level important. Here, however, we want to draw attention to systems where focusing only on the band structure is not enough. In particular, in the following, we will show examples where the form factor vanishes in crucial regions of the Brillouin zone, nullifying the large contribution from the DOS term.

\subsection{Flat band models with the $S$-matrix construction}

\label{sec: s-matrix}

We apply the pairing susceptibility approach to a class of bipartite flat-band models, constructed with the $S$-matrix method~\cite{calugaruGeneralConstructionTopological2022}. In a bipartite lattice with sublattices $L$ and $S$ such that $n_L > n_S$, where $n_L$ and $n_S$ are the numbers of orbitals in the large and small sublattice, respectively, there are no internal hoppings in the sublattices and the Fourier-transformed kinetic Hamiltonian is~\cite{calugaruGeneralConstructionTopological2022}
\begin{align}
	H_{\bm{k}} = \begin{pNiceMatrix}[first-row, last-col]
		n_L &  n_S\\
		0 & S_{\bm{k}} & n_L\\
		S_{\bm{k}}^\dagger & 0 & n_S\\
	\end{pNiceMatrix} \hspace{2pt}.
 \label{eq: H_kin_bipartite}
\end{align}
Here $S_{\bm{k}}$ is a $n_L \times n_S$ rectangular matrix which, due to the mismatched dimensions $n_L > n_S$, has at least $n_L - n_S$ zero eigenvalues for any $\bm{k}$, i.e. flat bands pinned at zero energy. The eigenvectors are 
$\ket{m_{\bm{k}}}^\text{flat} = \begin{pmatrix}
    \phi_{\bm{k}m} & 0
\end{pmatrix}^T$,
where $\phi_{\bm{k}m}$ is in the kernel of $S_{\bm{k}}^\dagger$.

In a model with no on-site interactions, the structure of $H_{\bm{k}}$ (Eq.~\ref{eq: H_kin_bipartite}) has striking implications. Consider an off-site interaction occurring between the same pairs of sites as the bipartite hoppings (typically nearest-neighbor pairs). Then, because of the projection to the flat band which is localized fully on sublattice $L$, the real-space order parameters within each sublattice vanish, and $\Delta_{\bm{k}}$ has the same bipartite structure as $H_{\bm{k}}$. Consider now the quantity $\braket{m_{\bm{k}+\qpdw}|\delta \Delta_\mu(\bm{k})|n_{\bm{k}-\qpdw}}$, appearing in the form factor of the pairing susceptibility (Eq.~\eqref{eq: pair_susceptibility}), when $m, n$ label such flat band states. We obtain
\begin{align}
    &\exval{m_{\bm{k}+\qpdw}|\delta \Delta_\mu(\bm{k})|n_{\bm{k}-\qpdw}} \nonumber\\
    &= \begin{pmatrix}
        \phi_{\bm{k} + \bm{q} m}^\dagger & 0
    \end{pmatrix} \begin{pmatrix}
        0 & \partial_{\Delta_\mu^R} \Delta_{\bm{k}}^\text{LS}\\
        \partial_{\Delta_\mu^R} \Delta_{\bm{k}}^\text{SL} & 0
    \end{pmatrix} \begin{pmatrix}
        \phi_{\bm{k} - \bm{q} n}\\
        0
    \end{pmatrix} = 0,
\end{align}
where LS and SL stand for the inter-sublattice pairing. This means that the intra-flat-band contribution to the pairing susceptibility vanishes, even when the flat band DOS diverges. 
Another consequence of the bipartite structure of the interaction is that also the quasi-particle spectrum 
has zero eigenvalues (App. \ref{sec: supp_bipartite_bdg_hamiltonian})~\footnote{Our work is inspired by the results of Minh Tam where he, as a summer trainee in our group in 2022, found that the nearest-neighbor pairing BdG spectrum of the Lieb lattice has flat bands.}.
In summary, in the case of nearest-neighbor pairing, flat bands are largely irrelevant for pairing in models constructed with the $S$-matrix method. However, nearest-neighbor pairing between the flat and the dispersive bands is still possible for large interactions, as well as nearest-neighbor pairing in non-bipartite lattice flat bands,  as we will show in the next section.

\begin{figure}
    \centering
    \includegraphics[width=\linewidth]{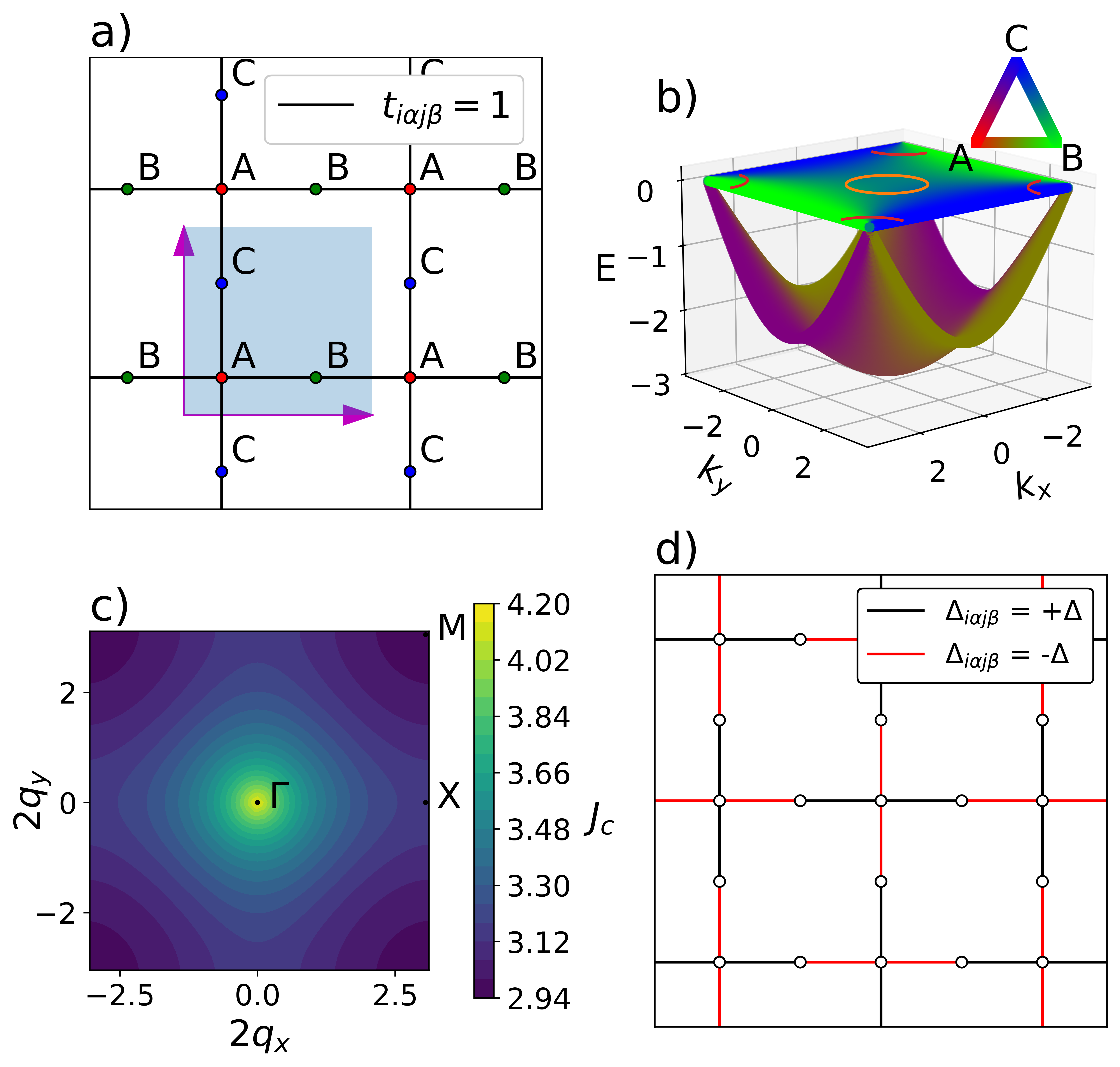}
    \caption{The Lieb lattice structure (\textbf{a}) and dispersion (\textbf{b}) showing a flat band and one of the two dispersive bands. The color indicates the orbital composition of each state. The orange and red circles highlight regions of large and small gradients in orbital composition near the flat band center and corners, respectively. Only the dispersive band with negative energies (bandwidth $2\sqrt{2}$) is shown; there is another, mirror image of it with positive energies (App. \ref{sec: supp_lieb_numerical_results} Fig.~\ref{fig:suppl_lieb_results}(e)). (\textbf{c}) The critical interaction strength $J_c$ as a function of the wave vector $2\qpdw$ with $T = 10^{-6}$. The minimum is found at the $\bm{M}$-point (with all four $\bm{M}$-points being equivalent), implying the formation of a PDW state. (\textbf{d}) The spatial structure of the PDW order parameter. A PDW with $2\qpdw = \bm{M}$ preserves the $C_{4v}$ symmetry of the lattice, and this order parameter structure belongs to the irreducible representation (irrep) $B_1$, when the origin is placed on one of the central $A$ orbitals.}
    \label{fig:lieb_results}
\end{figure}

\section{Application to example systems}

\subsection{Bipartite lattice example: the Lieb lattice}

We now focus on the Lieb lattice, see Fig.~\ref{fig:lieb_results}(a)-(b). In our Lieb and kagome lattice examples, we use a form of interaction~\cite{andersonResonatingValenceBond1987, black-schafferResonatingValenceBonds2007, baskaranResonatingValenceBond2017, julkuSuperfluidWeightBerezinskiiKosterlitzThouless2020} which is often used for describing 
nearest-neighbor singlet
pairing and is given by $\hat{H}_{\text{int}} = - \frac{J}{2} \sum_{\exval{i \alpha, j \beta}} \hod{i}{\alpha}{j}{\beta} \ho{i}{\alpha}{j}{\beta}$, where $\ho{i}{\alpha}{j}{\beta} = \co{i}{\alpha}{\uparrow} \co{j}{\beta}{\downarrow} - \co{i}{\alpha}{\downarrow} \co{j}{\beta}{\uparrow}$ such that $\ho{i}{\alpha}{j}{\beta}/\sqrt{2}$ is a singlet-pair annihilation operator. The sum is taken over nearest-neighbor pairs. This Hamiltonian corresponds to setting $A_\uparrow = 1/2, A_\downarrow = -1/2$ in Eq.~\eqref{eq: H_int}, accounting for an extra factor of two from restricting the sum to nearest neighbors. Since $\ho{i}{\alpha}{j}{\beta} = \ho{j}{\beta}{i}{\alpha}$, there is one independent singlet order parameter $\deltareal = -J \exval{\ho{i}{\alpha}{j}{\beta}}/2$ per nearest-neighbor pair, and the pairing matrix $\Delta_{\bm{k}}$ takes the form $[\Delta_{\bm{k}}]_{\alpha \beta} = -\sum_j \overline{\Delta}_{0 \alpha j \beta} e^{i \bm{k} \cdot (\bm{r}_{j \beta} - \bm{r}_{0 \alpha})}$. From now on, we work in units where $k_B = 1$, and also set the hopping amplitudes $t_{0 \alpha j \beta}^\sigma$ (our energy scale) to unity.

With the on-site Hubbard-$U$ interaction, any attractive interaction strength yields a uniform ($\qpdw = 0$) Bardeen-Cooper-Schrieffer (BCS) state for the Lieb lattice flat band~\cite{julkuGeometricOriginSuperfluidity2016, Huhtinen2022FlatBandSCQuantumMetric}. 
For the nearest-neighbor interactions we find, see Fig.~\ref{fig:lieb_results}(c), that $\qpdw = 0$ is not the ground state: the lowest critical interaction strength $J_c \approx 2.9$ is found at the $\bm{M}$-point $2 \qpdw = (\pi, \pi)^T$, corresponding to a PDW state. 
Why $2 \qpdw = \bm{M}$ gives the ground state is an intriguing interference effect arising from orbital-off-diagonal elements of the pairing susceptibility. The pairing is primarily interband, between dispersive band states near the Dirac point and flat band states near the $\Gamma$-point (Fig.~\ref{fig: supp_lieb_delta_band_k_resolved}(a) in App. \ref{sec: supp_lieb_numerical_results}). An analysis of the pairing for a circle of dispersive band momenta (angle $\theta$, radius $k_0$) around the Dirac point reveals that $\bm{q}$ is determined by maximizing the quantity ($n$ denotes the flat band) $\sqrt{[F^n_{BB}(\bm{q}) - F^n_{CC}(\qpdw)]^2 + 4 |F^n_{BC}(\bm{q})|^2}$ where 
    $F^n_{\alpha \beta}(\qpdw) = \frac{1}{2\pi} \int_0^{2\pi} f^n_\alpha(\theta, \qpdw)^* f^n_\beta(\theta, \qpdw) \intd{\theta}$ and $f^n_\alpha(\theta, \qpdw)$ are the orbital weights of the Bloch states (see App.~\ref{sec: supp_lieb_pdw_origin}). The maximum is reached when the 
Cauchy-Schwarz inequality 
   $ |F^n_{BC}(\bm{q})|^2 \leq F^n_{BB}(\bm{q}) F^n_{CC}(\bm{q})
$ is tight.
This occurs around the $\Gamma$-point (the orange circle in Fig.~\ref{fig:lieb_results}b) where $f^n_\alpha(\theta, \qpdw)$ are constant in $\theta$, therefore $2\bm{q}=\bm{M}$ gives the ground state. In contrast, around the Dirac point (magenta quadrants, pairing with $\bm{q}=0$) $F^n_{BC}(\bm{q})$ is close to zero. 
This shows that quantum geometry, i.e.~$\bm{k}$-dependence of the orbital composition of Bloch states, can also be detrimental to pairing even when it typically enhances superconductivity. 

We have confirmed that the PDW state is the ground state by minimizing the grand potential (App. \ref{sec: supp_lieb_numerical_results}). In real space, $2\qpdw = \bm{M}$ corresponds to the order parameters flipping their sign from one unit cell to the next; see Fig.~\ref{fig:lieb_results}(d). The PDW state spontaneously breaks the translational symmetry of the lattice. Note, however, that the critical interaction is very large, 2.9 times the hopping energy $t$, which is not typically the case when $J \sim t^2/U$ is derived perturbatively ($U$ is the repulsive on-site interaction), but can be achieved for suitable conditions~\cite{Han2020_HolsteinHubbard}. Much smaller PDW critical interactions can be found in the kagome lattice.

\begin{figure}
    \centering
    \includegraphics[width=\linewidth]{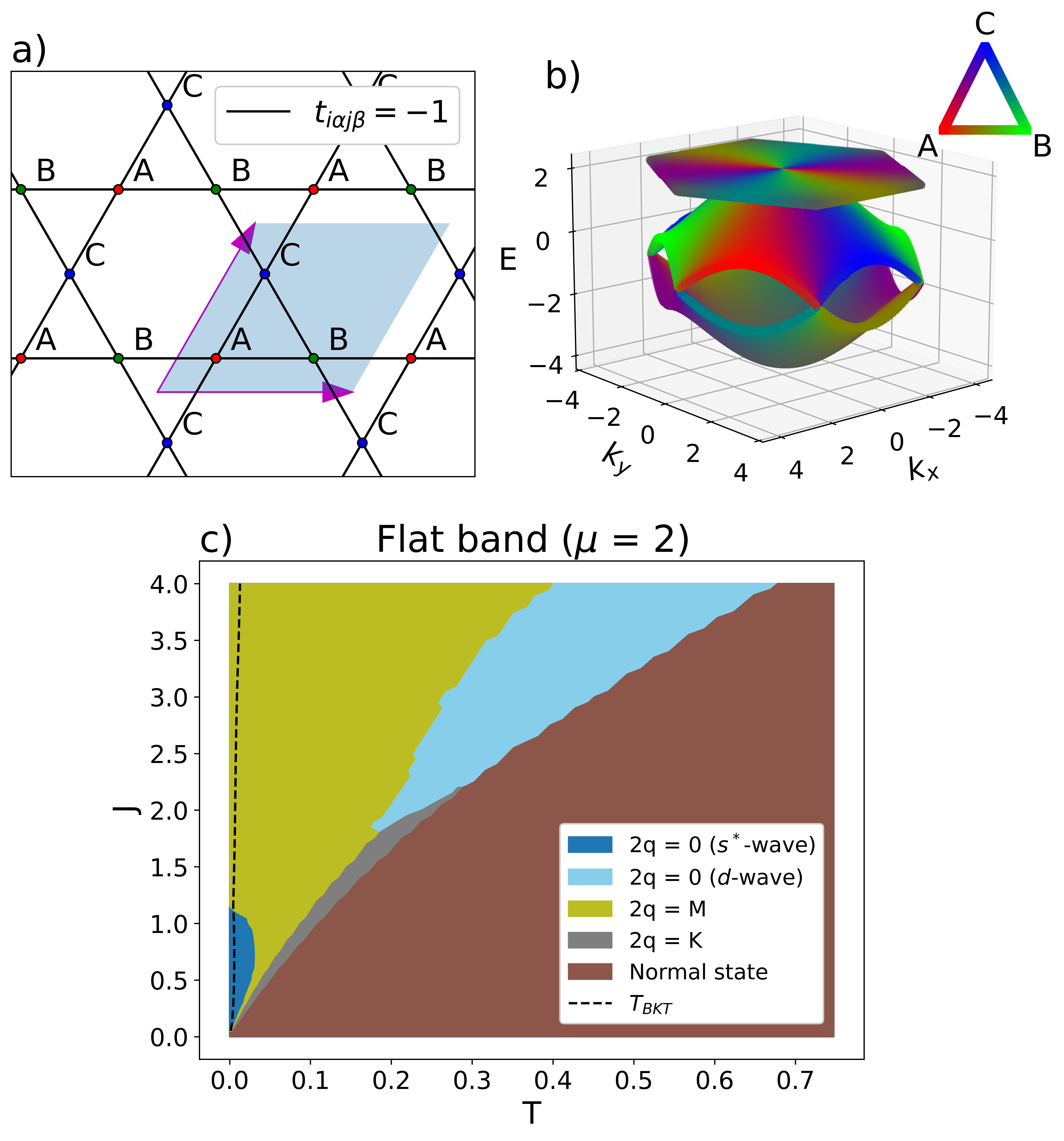}
    \caption{The structure (\textbf{a}) and dispersion (\textbf{b}) of the kagome lattice. The color indicates the orbital composition of each state. (\textbf{c}) A phase diagram for the flat band with two uniform BCS phases ($\qpdw = 0$, one with $s^*$-wave and another with $d$-wave pairing symmetry) and two different PDW phases ($2 \qpdw = \bm{M}, \bm{K})$. The black dashed line shows the BKT temperature.}
    \label{fig:kagome_fb_results}
\end{figure}

\subsection{Non-bipartite lattice example: kagome lattice}

\subsubsection{The flat band}

For a second example highlighting the importance of the Bloch state form factor in the pairing susceptibility, we turn to the kagome lattice, Figs.~\ref{fig:kagome_fb_results}(a)-(b). Now the mechanism leading to the suppression of pairing in bipartite lattice flat bands does not occur, and indeed we find a nearest-neighbor pairing ground state with $\qpdw = 0$ at the flat band ($\mu = 2$) for infinitesimally small values of $J$ at $T=0$, and for $J>1.1$ a PDW state with $2\qpdw = \bm{M}$.
At finite temperatures, however, $2\qpdw = \bm{K}$ PDW order appears too, as well as another $\qpdw = 0$ phase,
see Fig.~\ref{fig:kagome_fb_results}(c). The low-temperature $\qpdw = 0$ phase belongs to irrep $A_1$ of the point group $C_{6v}$, i.e. it has $s^*$-wave symmetry (fully homogeneous in real space). The high-temperature $\qpdw = 0$ phase instead belongs to the two-dimensional irrep $E_2$, and consists of both $d_{x^2-y^2}$ and $d_{xy}$-wave solutions, as well as combinations thereof (see Ref. \cite{wenSuperconductingPairingSymmetryKagome2022} for a detailed discussion of the pairing symmetries of the kagome lattice). The $2\qpdw = \bm{M}$ phase breaks the $C_{6v}$ symmetry into $C_{2v}$, and is a combination of multiple irreps (which is possible sufficiently far below $T_c$). Similarly, the $2\qpdw = \bm{K}$ phase breaks $C_{6v}$ into $C_{3v}$, and we find the solution to belong to irrep $A_1$ (homogeneous real-space pairing within one hexagon of the lattice, but different phases between hexagons due to non-zero $\qpdw$).

\subsubsection{The van Hove singularity at $\mu = 0$}

\begin{figure}
    \centering
    \includegraphics[width=\linewidth]{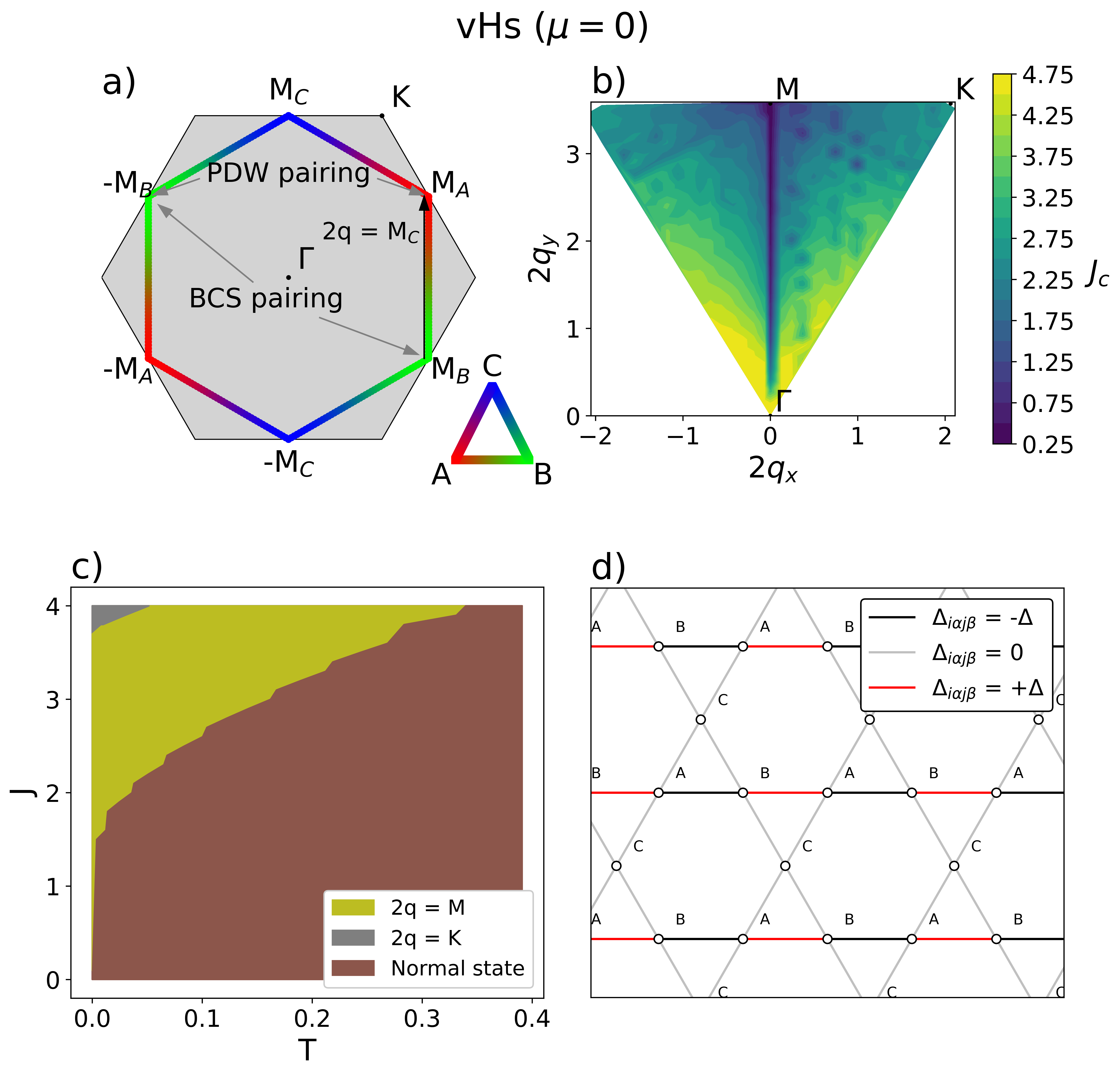}
    \caption{(\textbf{a}) The Fermi surface of the kagome lattice at $\mu = 0$ is a hexagon, with the vHs's at the corner points $\pm \bm{M}_\alpha$. The different $\bm{M}$-points are completely localized to orbital $\alpha$, shown in color. For a particle at $-\bm{M}_B$, typical $\qpdw = 0$ Cooper pairing with $\bm{M}_B$ is impossible with nearest-neighbor interactions due to localization to the same orbital. When $\qpdw = 2 \bm{M}_C$, pairs can instead be formed between $-\bm{M}_B$ and $\bm{M}_A$. (\textbf{b}) The critical interaction strength $J_c$ as a function of the PDW wave vector $2\qpdw$ with $T = 10^{-6}$. The minimum is found at the $\bm{M}$-point due to sublattice interference. Only one-sixth of the Brillouin zone is shown; the other regions are equivalent due to symmetry. (\textbf{c}) A phase diagram for the vHs. Here we see mainly a single PDW state with $2\qpdw = \bm{M}$, with an eventual transition to $2\qpdw = \bm{K}$ for very strong interactions. At $T=0$, the $2\qpdw = \bm{M}$ state exists for any positive $J$, but its critical temperature is very small for $J \lesssim 1.5$. The BKT temperature is zero, or at least vanishingly small throughout the $2\qpdw = \bm{M}$ phase. The $2\qpdw = \bm{M}$ phase breaks the $C_{6v}$ symmetry of the lattice to $C_{2v}$, and the solution belongs to irrep $A_1$. For the $2\qpdw = \bm{K}$ phase, the symmetry is broken to $C_{3v}$, and the solution belongs to the two-dimensional irrep $E$. (\textbf{d}) The real-space structure of the $2\qpdw = \bm{M}_C$ PDW state, in which case pairing occurs between orbitals $A$ and $B$. The other permutations are equivalent.
    }
    \label{fig:kagome_vhs_results}
\end{figure}

A second interesting point of the kagome spectrum is the vHs at $\mu=0$, which shows a strong dependence on the structure of the form factor. Fig.~\ref{fig:kagome_vhs_results}(a) illustrates the peculiarity of this point; the Fermi surface is a hexagon, with the vHs's at its corners at the $\bm{M}$-points. We denote the six different $\bm{M}$-points by $\pm \bm{M}_\alpha$, since the Fermi surface exhibits a phenomenon known as sublattice interference \cite{kieselSublatticeInterferenceKagome2012, wuSublatticeInterferencePromotes2023, linSublatticePolarizationDestructive2024}, characterized by the states at $\pm \bm{M}_\alpha$ being completely localized to orbital $\alpha$. As typical $\qpdw = 0$ Cooper pairing occurs between states at quasi-momenta $\bm{k}$ and $-\bm{k}$, this means that nearest-neighbor (i.e.~inter-orbital) pairing is inhibited between states at the vHs's, because the states are localized to the same orbital. Accordingly, with nearest-neighbor interactions, we find the ground state to be a PDW state with $2\qpdw = \pm \bm{M}_\alpha$, with $J_c$ close to zero (Fig. \ref{fig:kagome_vhs_results}(b). The appearance of the $2\qpdw = \pm \bm{M}_\alpha$ state is intuitive: with this $\qpdw$, Cooper pairs can form between two non-opposite $\bm{M}$-points which are localized to neighboring orbitals in real space, see Fig.~\ref{fig:kagome_vhs_results}(a). Since sublattice interference continues to energies away from vHs, see Fig.~\ref{fig:kagome_fb_results}(b), the same $2\qpdw = \pm \bm{M}_\alpha$ state preserves for a large range of $J$ and $T$, see Fig.~\ref{fig:kagome_vhs_results}(c)). These results are also reflected in the form factor of the pairing susceptibility (Eq.~\eqref{eq: pair_susceptibility}). Consider again the quantity $\exval{m_{\bm{k}+\qpdw}|\delta \Delta_\mu (\bm{k})|n_{\bm{k}-\qpdw}}$. When $\qpdw = 0$, the vHs's give terms like $\exval{\text{vHs}_{\pm \bm{M}_\alpha}|\delta \Delta_\mu (\bm{k})|\text
{vHs}_{\pm \bm{M}_\alpha}}$, which vanish because, due to sublattice interference, the Bloch states at the vHs's in the orbital basis are simply $\ket{\text{vHs}_{\pm \bm{M}_\alpha}} = \begin{pmatrix} 1 & 0 & 0 \end{pmatrix}^T$, $\begin{pmatrix} 0 & 1 & 0 \end{pmatrix}^T$, $\begin{pmatrix} 0 & 0 & 1 \end{pmatrix}^T$ for $\alpha = A, B, C$ respectively, and $\delta \Delta_\mu (\bm{k})$ is off-diagonal. In contrast, with $2 \qpdw = \pm \bm{M}_\alpha$ we obtain terms like $\exval{\text{vHs}_{\pm \bm{M}_\beta}|\delta \Delta_\mu (\bm{k})|\text{vHs}_{\pm \bm{M}_\gamma}}$ with $\beta \neq \gamma$, which are non-zero. Further information about the pairing via band-resolved pictures is given in App. \ref{sec: supp_kagome_results} \cref{fig: supp_lieb_delta_band_k_resolved,fig: suppl_kagome_M_delta_band_k_resolved,fig: suppl_kagome_G_K_delta_band_k_resolved}.

Various types of interactions on the kagome lattice at van Hove filling have been predicted to favor $2\qpdw = \bm{M}$ charge density wave (CDW) and PDW processes, using renormalization group methods~\cite{wangCompetingElectronicOrders2013, kieselSublatticeInterferenceKagome2012, wuSublatticeInterferencePromotes2023}, bare and random phase approximation (RPA) susceptibility calculations~\cite{fuExoticChargeDensity2025, dongLoopcurrentChargeDensity2023}, and mean-field theory~\cite{dongLoopcurrentChargeDensity2023}. For the interactions we consider, we find that although the susceptibility predicts the existence of a $2\qpdw = \bm{M}$ PDW order at the vHs, the PDW state is unstable to fluctuations of the order parameter phase, as will be discussed in the next section.

\begin{figure}
    \centering
    \includegraphics[width=\linewidth]{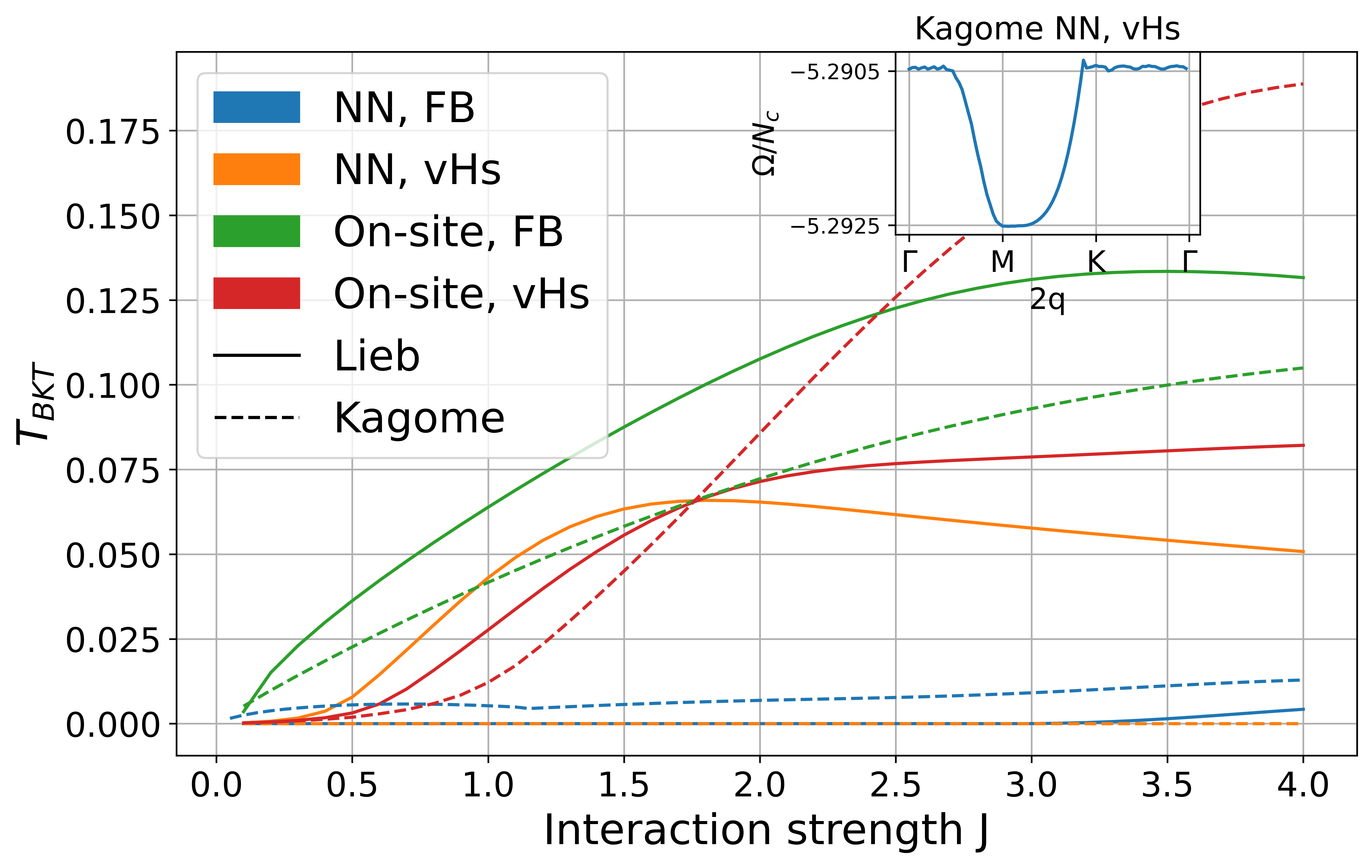}
    \caption{BKT temperatures for the Lieb and kagome lattices with both on-site (for which $J$ should be understood as the Hubbard $U$) and nearest-neighbor (NN) interactions, and with the chemical potential set to the flat band (FB) and the vHs. The inset shows the flatness of the grand potential $\Omega$ as a function of $2\qpdw$ near its minimum at $\bm{M}$ for the kagome vHs with NN interactions for $J = 2.5$ and $T = 10^{-6}$, indicating a vanishing superfluid weight and thus $T_{\text{BKT}}$ (the dashed orange line).}
    \label{fig: bkt_temperatures}
\end{figure}

\section{Superfluid weight}

The superfluid weight $D_s$ is a measure of the stiffness against phase fluctuations of the order parameter and thereby surpasses the mean-field treatment. It determines whether the ground state can support supercurrent, and in 2D systems gives the BKT transition temperature via~\cite{berezinskyDestructionLongRange1971, kosterlitzOrderingMetastabilityPhase1973, nelsonUniversalJumpSuperfluid1977} $T_{\text{BKT}} = \frac{\pi}{8} \sqrt{\det(D_s(T_{\text{BKT}}))}$. We define $D_s$ in terms of the grand potential as
\begin{align}
    [D_s]_{ij} = \frac{1}{V} \frac{\text{d}^2 \Omega(\bm{q}')}{\text{d} q'_i \text{d} q'_j} \bigg |_{\bm{q}' = \qpdw},
    \label{eq: sfw}
\end{align}
where $V$ is the volume of the system (area in 2D). This definition naturally extends the typical BCS superfluid weight definition to PDW states simply by evaluating the derivative at the PDW wave vector $\bm{q}' = \qpdw$ instead of at $\bm{q}' = 0$. In App. \ref{sec: supp_sfw_lin_response}, we also show how this extension agrees with the alternative but equivalent definition of the superfluid weight based on linear response theory. In previous literature, PDW states are sometimes associated with the superfluid weight of a conventional $\bm{q} = 0$ BCS state becoming non-positive-semidefinite~\cite{jiangPairDensityWaves2023, chenPairDensityWave2023} indicating that $\bm{q} = 0$ is no longer a minimum of the grand potential, and the system becomes unstable towards a PDW state. However, the definition of the superfluid weight we are using here is always positive-semidefinite, since $\qpdw$ is by definition a minimum of the grand potential. Thus this definition properly measures the stability of the PDW state.

We derive the superfluid weight formula for the nearest-neighbor pairing (PDW or $\bm{q}=0$) superconducting orders, while properly taking into account the implicit dependence of the order parameters on $\bm{q}$~\cite{Huhtinen2022FlatBandSCQuantumMetric,Tam2024GeomIndependence}, 
and then evaluate the superfluid weight numerically (details in App. \ref{sec: supp_sfw}). In figure \ref{fig: bkt_temperatures}, we show the BKT temperatures obtained from the superfluid weight results for the Lieb and kagome lattices, for both nearest-neighbor and on-site ($J$ should be understood as the Hubbard-$U$) interactions. We find that while the PDW states with the nearest-neighbor interaction can give non-zero superfluid weights, the BKT temperatures are much smaller than for the corresponding $\bm{q} = 0$ states with on-site interactions. This is the case for the flat bands of both Lieb and kagome lattices (solid and dashed blue), which means only the low-temperature $\bm{q} = 0$ and $2\bm{q} = \bm{M}$ phases are stable in the kagome flat band phase diagram (figure~\ref{fig:kagome_fb_results}(c)). At the kagome vHs (dashed orange), the BKT temperature vanishes due to instability to phase fluctuations of the order parameter, as seen from the flatness of the grand potential minimum in the inset of Fig.~\ref{fig: bkt_temperatures}. This can be understood to arise from the shape of the Fermi surface at the vHs, see App. \ref{sec: supp_kagome_results} Fig.~\ref{fig:suppl_kagome_vhs_sfw_schematic}. However, for the Lieb lattice with nearest-neighbor interactions, we find that moving away from the flat band (solid blue) to the vHs (solid orange) significantly enhances the BKT temperature, making it comparable to the on-site case (solid red). The ground state for the vHs is a $\qpdw = 0$ state (App. \ref{sec: supp_lieb_numerical_results}). This further demonstrates how the flat band becomes nearly irrelevant with non-local interactions due to the bipartite structure of the lattice. Interestingly, in the kagome lattice with on-site interactions, the vHs (dashed red) gives a larger BKT temperature than the flat band (dashed green) when the interaction strength $J \gtrsim 1.8$, even when its order parameter is smaller (Fig.~\ref{fig:suppl_deltas_sfws} in App. \ref{sec: supp_kagome_results}). 

\section{Conclusions}

We expressed the pairing susceptibility in a form that explicitly shows the interplay between the orbital compositions of the eigenstates (quantum geometry) and of the order parameter. This allows, for instance, to easily see that nearest-neighbor pairing is forbidden for the flat band states of bipartite lattices. In the Lieb lattice flat band, we found a PDW with a vector $2\bm{q}=\bm{M}$ originating from a subtle interference effect related to the orbital composition of the flat band states: this is an example of that quantum geometry can be detrimental to (in this case to $2\bm{q}=0$) pairing. In the kagome lattice flat band, superconductivity is possible at infinitesimally small interaction, but the BKT temperature is suppressed compared to on-site s-wave pairing. At the kagome vHs, although pairing susceptibility predicts a PDW order, we find its superfluid weight zero -- this suggests it is essential to determine the superfluid weight for previous and future predictions of PDW orders in kagome van Hove and similar singularities. High DOS typically
enhances pairing susceptibility, therefore DOS singularities are considered as a potential way to enhance superconductivity. However, our results show that a careful analysis
of the eigenstate structure is crucial since nearest-neighbor pairing is sensitive to the orbital composition of the eigenstates. Further, our results emphasize the importance of calculating the superfluid weight (stiffness) at DOS singularities, as we found several
examples where it was extremely small or even zero, even when pairing susceptibility predicted a sizeable superconducting gap. The importance of the eigenstate structure and phase stiffness can be relevant also when understanding the competition between superconductivity, magnetism, and charge density waves~\cite{Hofmann2023SC_CWD} in flat bands and different types of van Hove singularities~\cite{Classen2024, scammellChiralExcitonicOrder2023}, an important area of future study.


\section*{Acknowledgments}

We thank Minh Tam for the preliminary work that inspired us to start this project. We thank Andrei Bernevig, Jonah Herzog-Arbeitman, Zhaoyu Han, Kukka-Emilia Huhtinen, and Gabriel Topp for useful discussions. We acknowledge the computational resources provided by the Aalto Science-IT project. This work was supported by the Research Council of Finland under project number 339313, by Jane and Aatos Erkko Foundation, Keele Foundation, Magnus Ehrnrooth Foundation, and Klaus Tschira Foundation as part of the SuperC collaboration, and by a grant from the Simons Foundation (SFI-MPS-NFS-00006741-12, P.T.) in the Simons Collaboration on New Frontiers in Superconductivity.






\appendix

\section{Derivation of the BdG Hamiltonian}

\label{sec: supp_bdg_hamiltonian}

Here, we derive the BdG form of the Hamiltonian, Eqs.~\eqref{eq: H_bdg}-\eqref{eq: delta_k} in the main text. After a mean-field decomposition, the Hamiltonian reads

\begin{align}
    \hat{H} &= \sum_{i \alpha, j\beta} \sum_\sigma t_{i \alpha j \beta}^\sigma \cod{i}{\alpha}{\sigma} \co{j}{\beta}{\sigma} - \mu \sum_{i \alpha \sigma} \cod{i}{\alpha}{\sigma} \co{i}{\alpha}{\sigma}\\
    &+ \sum_{i \alpha, j \beta} \bigg(\deltareal^* \bo{i}{\alpha}{j}{\beta} + \text{H.c} + |\deltareal|^2 / J_{i \alpha j \beta}\bigg), \nonumber\\
    \bo{i}{\alpha}{j}{\beta} &= \sum_\sigma A_\sigma \co{i}{\alpha}{\sigma} \co{j}{\beta}{-\sigma}, \label{eq_supp: b_op_definition}\\
    \deltareal &= -J_{i \alpha j \beta} \exval{\bo{i}{\alpha}{j}{\beta}}.
\end{align}
Next, we introduce a Fourier transform $\co{i}{\alpha}{\sigma} = (1/\sqrt{N_c}) \sum_{\bm{k}} e^{i \bm{k} \cdot \bm{r}_{i \alpha}} \co{\bm{k}}{\alpha}{\sigma}$, where $N_c$ is the number of unit cells, and $\bm{k}$ is the momentum with the sum taken over the first Brillouin zone. The kinetic Hamiltonian and the chemical potential term, i.e. the first two terms of the Hamiltonian above, transform in a standard way, giving

\begin{align}
    \sum_{i \alpha, j\beta} \sum_\sigma t_{i \alpha j \beta}^\sigma \cod{i}{\alpha}{\sigma} \co{j}{\beta}{\sigma} &= \sum_{\bm{k} \alpha \beta} \sum_\sigma [H_{\bm{k}}^\sigma]_{\alpha \beta} \cod{\bm{k}}{\alpha}{\sigma} \co{\bm{k}}{\beta}{\sigma},\\
    \mu \sum_{i \alpha \sigma} \cod{i}{\alpha}{\sigma} \co{i}{\alpha}{\sigma} &= \mu \sum_{\bm{k} \alpha \sigma} \cod{\bm{k}}{\alpha}{\sigma} \co{\bm{k}}{\alpha}{\sigma},\\
    [H_{\bm{k}}^\sigma]_{\alpha \beta} &= \sum_j t_{0 \alpha j \beta}^\sigma e^{i \bm{k} \cdot \bm{r}^\Delta_{0 \alpha j \beta}},\\
    \bm{r}^\Delta_{i \alpha j \beta} &\equiv \bm{r}_{j \beta} - \bm{r}_{i \alpha}.
\end{align}
Applying the Fourier transform to the interaction term gives

\begin{widetext}
\begin{align}
    \sum_{i \alpha, j \beta} \deltareal^* \bo{i}{\alpha}{j}{\beta} &= - \frac{1}{N_c} \sum_{i \alpha, j \beta} \sum_{\bm{k} \bm{k}' \sigma} A_\sigma \Delta^*_{i \alpha j \beta} e^{i (\bm{k} \cdot \bm{r}_{i \alpha} + \bm{k}' \cdot \bm{r}_{j \beta})} \co{\bm{k}}{\alpha}{\sigma} \co{\bm{k}'}{\beta}{-\sigma}\\
    &= \frac{1}{N_c} \sum_{i \alpha, j \beta} \sum_{\bm{k} \bm{k}' \sigma} A_\sigma \overline{\Delta}^*_{0 \alpha j \beta} e^{-i 2 \qpdw \cdot (\bm{r}_{i \alpha} + \bm{r}_{(i+j)\beta})/2} e^{i (\bm{k} \cdot \bm{r}_{i \alpha} + \bm{k}' \cdot \bm{r}_{(i+j) \beta})} \co{\bm{k}}{\alpha}{\sigma} \co{\bm{k}'}{\beta}{-\sigma}\\
    &= \sum_{j \alpha \beta \bm{k} \sigma} A_\sigma \overline{\Delta}^*_{0 \alpha j \beta} e^{-i(\bm{k} - \qpdw) \cdot \bm{r}^\Delta_{0 \alpha j \beta}} \co{\bm{k}}{\alpha}{\sigma} \co{-\bm{k}+2\qpdw}{\beta}{-\sigma}.
\end{align}
Here we have made an index shift $j \rightarrow j + i$, substituted the definition of $\overline{\Delta}_{0 \alpha j \beta}$ (Eq.~\eqref{eq: delta_real_formulation} in the main text), and used the Fourier transform property $\sum_i e^{i(-2 \qpdw + \bm{k} + \bm{k}') \cdot \bm{R}_{i}} = N_c \delta_{\bm{k}',-\bm{k}+2\qpdw}$, where $\delta$ is the Kronecker delta. Continuing with another index shift $\bm{k} \rightarrow -\bm{k} + \qpdw$ and separating the two spin species, we have
\begin{align}
    \sum_{i \alpha, j \beta} \deltareal^* \bo{i}{\alpha}{j}{\beta} &= \sum_{j \alpha \beta \bm{k}} \bigg( A_\uparrow \overline{\Delta}^*_{0 \alpha j \beta} e^{i\bm{k} \cdot \bm{r}^\Delta_{0 \alpha j \beta}} \co{-\bm{k}+\qpdw}{\alpha}{\uparrow} \co{\bm{k}+\qpdw}{\beta}{\downarrow} + A_\downarrow \overline{\Delta}^*_{0 \alpha j \beta} e^{i\bm{k} \cdot \bm{r}^\Delta_{0 \alpha j \beta}} \co{-\bm{k}+\qpdw}{\alpha}{\downarrow} \co{\bm{k}+\qpdw}{\beta}{\uparrow} \bigg)\\
    &= \sum_{j \alpha \beta \bm{k}} \bigg( - A_\uparrow \overline{\Delta}^*_{0 \beta j \alpha} e^{-i\bm{k} \cdot \bm{r}^\Delta_{0 \beta j \alpha}} \co{-\bm{k}+\qpdw}{\alpha}{\downarrow} \co{\bm{k}+\qpdw}{\beta}{\uparrow} + A_\downarrow \overline{\Delta}^*_{0 \alpha j \beta} e^{i\bm{k} \cdot \bm{r}^\Delta_{0 \alpha j \beta}} \co{-\bm{k}+\qpdw}{\alpha}{\downarrow} \co{\bm{k}+\qpdw}{\beta}{\uparrow} \bigg)\\
    &= \sum_{\bm{k} \alpha \beta} [\Delta_{\bm{k}}]^*_{\beta \alpha} \co{-\bm{k}+\qpdw}{\alpha}{\downarrow} \co{\bm{k}+\qpdw}{\beta}{\uparrow},
\end{align}
\end{widetext}
where $[\Delta_{\bm{k}}]^*_{\beta \alpha} \equiv \sum_j A_\downarrow \overline{\Delta}^*_{0 \alpha j \beta} e^{i\bm{k} \cdot \bm{r}^\Delta_{0 \alpha j \beta}} - A_\uparrow \overline{\Delta}^*_{0 \beta j \alpha} e^{-i\bm{k} \cdot \bm{r}^\Delta_{0 \beta j \alpha}}$. To get the second line, we have used fermionic anticommutation relations, swapped $\alpha \leftrightarrow \beta$, and shifted $\bm{k} \rightarrow - \bm{k}$ on the first term of the right-hand side of the first line.


With these results, we can write the Hamiltonian in BdG form as

\begin{align}
   \hat{H} &= \sum_{\bm{k}} \cdspinor{\bm{k}} H_{\text{BdG}}(\bm{k}) \cspinor{\bm{k}} + N_c \sum_{j \alpha \beta} \frac{|\overline{\Delta}_{0 \alpha j \beta}|^2}{J_{0 \alpha j \beta}},\\
   H_{\text{BdG}}(\bm{k}) &= \begin{pmatrix}
       H^\uparrow_{\bm{k} + \qpdw} - \mu \bm{1} & \Delta_{\bm{k}}\\
       \Delta^\dagger_{\bm{k}} & - (H^\downarrow_{-\bm{k} + \qpdw})^* + \mu \bm{1}
   \end{pmatrix}, \label{eq_supp: H_bdg}\\
   [\Delta_{\bm{k}}]_{\alpha \beta} &= \sum_j \bigg( A_\downarrow \overline{\Delta}_{0 \beta j \alpha} e^{-i \bm{k} \cdot \bm{r}^\Delta_{0 \beta j \alpha}} \nonumber \\ &\qquad \qquad- A_\uparrow \overline{\Delta}_{0 \alpha j \beta} e^{i \bm{k} \cdot \bm{r}^\Delta_{0 \alpha j \beta}} \bigg),
   \label{eq_supp: delta_k}\\
   \cspinor{\bm{k}} &= (\co{\bm{k} + \qpdw}{\alpha=1}{\uparrow}, \dots, \co{\bm{k} + \qpdw}{\alpha=n_{\text{orb}}}{\uparrow}, \nonumber \\ &\qquad \cod{-\bm{k} + \qpdw}{\alpha=1}{\downarrow}, \dots, \cod{-\bm{k} + \qpdw}{\alpha=n_{\text{orb}}}{\downarrow})^T,
\end{align}
where constant terms that depend on neither operators nor the order parameters have been dropped and $n_{\text{orb}}$ is the number of orbitals. From now on, we will assume time-reversal symmetry (TRS), which implies $(H^\downarrow_{-\bm{k} + \qpdw})^* = H^\uparrow_{\bm{k} - \qpdw} \equiv H_{\bm{k} - \qpdw}$, allowing us to drop the spin indices.

In this work, we are especially interested in the structure of the pairing matrix $\Delta_{\bm{k}}$. For the standard on-site Hubbard-$U$ interaction, it reduces to a simple $\bm{k}$-independent diagonal matrix, $\Delta_{\bm{k}} = \text{diag}(\overline{\Delta}_{0 \alpha 0 \alpha})$, but as is evident from the above form, in general, it can be non-diagonal and $\bm{k}$-dependent. Often, the pairing matrix can be simplified further. For example, if $A_{\downarrow} = - A_\uparrow$, we have $\bo{i}{\alpha}{j}{\beta} = \bo{j}{\beta}{i}{\alpha}$ and thus $\Delta_{i \alpha j \beta} = \Delta_{j \beta i \alpha}$ by fermionic anticommutation relations, leading to

\begin{align}
    [\Delta_{\bm{k}}]_{\alpha \beta} &= - A_\uparrow \sum_j \bigg( \overline{\Delta}_{0 \beta j \alpha} e^{-i\bm{k} \cdot \bm{r}^\Delta_{0 \beta j \alpha}} + \overline{\Delta}_{0 \alpha j \beta} e^{i\bm{k} \cdot \bm{r}^\Delta_{0 \alpha j \beta}} \bigg)\\
    &= - A_\uparrow \sum_j \bigg( \overline{\Delta}_{j \beta 0 \alpha} e^{-i\bm{k} \cdot \bm{r}^\Delta_{j \beta 0 \alpha}} + \overline{\Delta}_{0 \alpha j \beta} e^{i\bm{k} \cdot \bm{r}^\Delta_{0 \alpha j \beta}} \bigg)\\
    &= - 2 A_\uparrow \sum_j \overline{\Delta}_{0 \alpha j \beta} e^{i\bm{k} \cdot \bm{r}^\Delta_{0 \alpha j \beta}}.
\end{align}
Here, we first transform $j \rightarrow -j$, then use $\overline{\Delta}_{0 \beta -j \alpha} = \overline{\Delta}_{j \beta 0 \alpha} = \overline{\Delta}_{0 \alpha j \beta}$ and $\bm{r}^\Delta_{0 \beta -j \alpha} = \bm{r}^\Delta_{j \beta 0 \alpha} = -\bm{r}^\Delta_{0 \alpha j \beta}$ on the first part of the sum. The nearest-neighbor interaction that was used for the numerical results of this work is a special case of this result with $A_\uparrow = 1/2$.

\section{Bipartite Form of the BdG Hamiltonian}

\label{sec: supp_bipartite_bdg_hamiltonian}

For the type of bipartite lattice discussed in Sec. \ref{sec: s-matrix} of the main text, i.e. one with sublattices $L$ and $S$ such that the number of orbitals $n_L > n_S$, and an interaction with a similar bipartite structure, the order parameter matrix $\Delta_{\bm{k}}$ is of the form

\begin{align}
   \Delta_{\bm{k}} = \begin{pNiceMatrix}[first-row, last-col]
       n_L & n_S\\
       0 & \Delta_{\bm{k}}^\text{LS} & n_L\\
       \Delta_{\bm{k}}^\text{SL} & 0 & n_S
   \end{pNiceMatrix} \hspace{2pt},
   \label{eq: delta_k_bipartite}
\end{align}
where LS and SL stand for the inter-sublattice pairing, and $n_L$ and $n_S$ refer here to the size of the matrix. Plugging in this and the bipartite form of $H_{\bm{k}}$, Eq.~\eqref{eq: H_kin_bipartite} in the main text, to the BdG-Hamiltonian, Eq.~\eqref{eq_supp: H_bdg} yields
\begin{align}
	H_{\text{BdG}}(\bm{k}) = \begin{pNiceMatrix}[first-row, last-col]
		n_L &  n_S & n_L &  n_S\\
			0 & S_{\bm{k}+\qpdw} & 0 & \Delta^{\text{UR}}_{\bm{k}} & n_L\\
			S_{\bm{k}+\qpdw}^\dagger & 0 & \Delta^{\text{LL}}_{\bm{k}} & 0 & n_S\\
			0 & (\Delta^{\text{LL}}_{\bm{k}})^\dagger & 0 & -S_{\bm{k}-\qpdw} & n_L\\
			(\Delta^{\text{UR}}_{\bm{k}})^\dagger & 0 & -S_{\bm{k}-\qpdw}^\dagger & 0 & n_S\\
	\end{pNiceMatrix},
\end{align}
where the chemical potential has been set to zero, i.e. to the flat band(s), and TRS has been assumed. Permuting the second and third rows and columns of this matrix (which is a unitary transformation) results in a bipartite form for the whole BdG Hamiltonian:

\begin{align}
	\tilde{H}_{\text{BdG}}(\bm{k}) = \begin{pNiceMatrix}[first-row, last-col]
		n_L &  n_L & n_S &  n_S\\
			0 & 0 & S_{\bm{k}+\qpdw} & \Delta^{\text{UR}}_{\bm{k}} & n_L\\
			0 & 0 & (\Delta^{\text{LL}}_{\bm{k}})^\dagger  & -S_{\bm{k}-\qpdw} & n_L\\
			S_{\bm{k}+\qpdw}^\dagger & \Delta^{\text{LL}}_{\bm{k}} & 0 & 0 & n_S\\
			(\Delta^{\text{UR}}_{\bm{k}})^\dagger & -S_{\bm{k}-\qpdw}^\dagger & 0 & 0 & n_S\\
	\end{pNiceMatrix}.
   \label{eq: H_bdg_bipartite}
\end{align}
This means that with this setup, the BdG spectrum has at least $2(n_L - n_S)$ zero eigenvalues for each $\bm{k}$, i.e. it inherits the flat band(s) from the single-particle spectrum as discussed in the main text.

\section{Grand Potential and its Derivatives}

\label{sec: supp_grand_potential}

The grand potential $\Omega$ of the system is given by

\begin{align}
   \Omega &= - k_B T \sum_{\bm{k} a} \ln (1 + e^{-E_{\bm{k}a}/ k_B T} ) + N_c \sum_{j \alpha \beta} \frac{|\overline{\Delta}_{0 \alpha j \beta}|^2}{J_{0 \alpha j \beta}},
   \label{eq_supp: omega}
\end{align}
where $E_{\bm{k}a}$ are the $2 n_{\text{orb}}$ eigenvalues of the BdG Hamiltonian $H_{\text{BdG}}(\bm{k})$, the sum over $\bm{k}$ is taken over the first Brillouin zone, and constants that do not depend on either operators or the order parameters have been ignored.


We need to take various second derivatives of the grand potential to calculate the pairing susceptibility as well as the superfluid weight. For that purpose, here we derive a generic expression for such derivatives. Straight-forward differentiation of the first term in Eq.~\eqref{eq_supp: omega} yields
\begin{align}
    &\partial_z \partial_{z'} \bigg( - k_B T \sum_{\bm{k} a} \ln (1 + e^{-E_{\bm{k}a}/ k_B T} ) \bigg) \nonumber\\ = &\sum_{\bm{k} a} n_F'(E_{\bm{k}a}) \partial_z E_{\bm{k}a} \partial_{z'} E_{\bm{k}a} + \sum_{\bm{k} a} n_F(E_{\bm{k}a}) \partial_z \partial_{z'} E_{\bm{k}a},
    \label{eq_supp: omega_hessian_intermediate_result}
\end{align}
where $n_F$ and $n_F'$ are the Fermi-Dirac distribution and its derivative, respectively, and $z$ and $z'$ are generic parameters that the matrix $H_{\text{BdG}}(\bm{k})$ depends on. To continue, we need to take derivatives of the BdG energies $E_{\bm{k}a}$. We can do this without knowing the explicit form of the energies by using the Hellmann-Feynman theorem:

\begin{widetext}
\begin{align}
    \partial_z E_{\bm{k}a} &= \braket{\phi_{\bm{k}a} | \partial_z H_{\text{BdG}}(\bm{k})|\phi_{\bm{k}a}},
    \label{eq_supp: dx_E}\\
    \partial_z \partial_{z'} E_{\bm{k}a} &= \braket{\phi_{\bm{k}a} | \partial_z \partial_{z'} H_{\text{BdG}}(\bm{k})|\phi_{\bm{k}a}} + \sum_{b \neq a} \frac{1}{E_{\bm{k}a} - E_{\bm{k}b}} \bigg( \braket{\phi_{\bm{k}a} | \partial_z H_{\text{BdG}}(\bm{k}) | \phi_{\bm{k}b}} \braket{\phi_{\bm{k}b} | \partial_{z'} H_{\text{BdG}}(\bm{k}) | \phi_{\bm{k}a}} + \text{H.c.} \bigg).
    \label{eq_supp: dxdy_E}
\end{align}
We denote the BdG quasi-particle states with $\ket{\phi_{\bm{k}a}}$ so that $H_{\text{BdG}}(\bm{k}) \ket{\phi_{\bm{k}a}} = E_{\bm{k}a} \ket{\phi_{\bm{k}a}}$. Substituting these results to Eq.~\eqref{eq_supp: omega_hessian_intermediate_result}, we get the following expression for a generic second derivative of the grand potential

\begin{equation}
\begin{aligned}
    \partial_z \partial_{z'} \Omega &= \sum_{\bm{k} a} \bigg( n_F'(E_{\bm{k}a}) \braket{\phi_{\bm{k}a} | \partial_z H_{\text{BdG}}(\bm{k})|\phi_{\bm{k}a}} \braket{\phi_{\bm{k}a} | \partial_{z'} H_{\text{BdG}}(\bm{k})|\phi_{\bm{k}a}} + n_F(E_{\bm{k}a}) \braket{\phi_{\bm{k}a} | \partial_z \partial_{z'} H_{\text{BdG}}(\bm{k})|\phi_{\bm{k}a}} \bigg)\\
    &+ \sum_{\bm{k} ab, a \neq b} \frac{n_F(E_{\bm{k}a}) - n_F(E_{\bm{k}b})}{E_{\bm{k}a} - E_{\bm{k}b}} \Re \bigg( \braket{\phi_{\bm{k}a} | \partial_z H_{\text{BdG}}(\bm{k}) | \phi_{\bm{k}b}} \braket{\phi_{\bm{k}b} | \partial_{z'} H_{\text{BdG}}(\bm{k}) | \phi_{\bm{k}a}} \bigg)\\
    &+ \partial_z \partial_{z'} \bigg( N_c \sum_{j \alpha \beta} \frac{|\overline{\Delta}_{0 \alpha j \beta}|^2}{J_{0 \alpha j \beta}} \bigg).
\end{aligned}
\label{eq_supp: omega_derivative_general}
\end{equation}
\end{widetext}
The summation in the second row has been symmetrized using $\sum_{ab} f(a, b) = \frac{1}{2} \sum_{ab} (f(a,b) + f(b,a))$. The advantage of this expression is that all of the derivatives have been shifted to the BdG Hamiltonian, and these are often available analytically, while for direct differentiation of the BdG energies numerical methods would typically have to be used.

\section{Full Derivation of the Pairing Susceptibility}

\label{sec: supp_pairing_susceptibility}

Calculating the Hessian of the grand potential, $\omegahessian \equiv \text{Hess}(\Omega) |_{\Vec{\Delta} = 0}$, amounts to taking derivatives of the grand potential with respect to both the real and imaginary parts of the order parameters, and evaluating these derivatives at $\Vec{\Delta} = 0$. Thus, we can directly apply Eq.~\eqref{eq_supp: omega_derivative_general}.

At this point, one should be careful with the proper definition of $\Vec{\Delta}$. The vector $\Vec{\Delta}$ should contain all \textit{independent} order parameters, which may not be the full set of $\overline{\Delta}_{0 \alpha j \beta}$ with all possible $j, \alpha, \beta$. For example, as mentioned in App. \ref{sec: supp_bdg_hamiltonian}, if $A_\downarrow = - A_\uparrow$ in the definition of the interaction, Eq.~\eqref{eq_supp: b_op_definition} (which holds for the NN interaction used in this work), we have $\Delta_{i \alpha j \beta} = \Delta_{j \beta i \alpha}$, which cuts the number of degrees of freedom in half (for the off-site portion of the interaction) to one per unique pair of sites $0 \alpha$ and $j \beta$. Here, we consider two possibilities: either all of the $\overline{\Delta}_{0 \alpha j \beta}$ are independent so that the degrees of freedom are the ordered pairs $0 \alpha, j \beta$ (for which $J_{0 \alpha j \beta} \neq 0$), or then $\Delta_{i \alpha j \beta} = \Delta_{j \beta i \alpha}$, in which case the degrees of freedom are the set of unordered pairs $0 \alpha, j \beta$. With slight modifications, the following derivation can be extended to other types of dependencies between the order parameters as well. Note that for the on-site portion of the interaction (pairs $0 \alpha, 0 \alpha$), the distinction between ordered and unordered pair is unnecessary. For ease of notation, we denote an element of $\Vec{\Delta}$ by $\Delta_\mu$, where $\{\mu\}$ is the appropriate set of degrees of freedom.

Since we are interested in $\Vec{\Delta} = 0$, we take advantage of the knowledge of the normal state properties of the system: the energies $E_{\bm{k}a}$ and eigenstates $\ket{\phi_{\bm{k}a}}$ reduce to the single-particle energies $\epsilon_{\bm{k}m}$ and Bloch states $\ket{m_{\bm{k}}}$. Specifically, when $\Vec{\Delta} = 0$ we have a pair of quasi-particle states for each Bloch state, and thus can use the band index $m$ to label them:

\begin{align}
    E_{\bm{k}m}^\pm &= \pm (\epsilon_{\bm{k} \pm \qpdw m} - \mu) \equiv \pm \xi_{\bm{k} \pm \qpdw m},\\
    \ket{\phi^+_{\bm{k}m}} &= \colvec{2}{\ket{m_{\bm{k} + \qpdw}}}{0}, \label{eq_supp: phi_plus}\\
    \ket{\phi^-_{\bm{k}m}} &= \colvec{2}{0}{\ket{m_{\bm{k} - \qpdw}}}. \label{eq_supp: phi_minus}
\end{align}
TRS has been assumed in the above result. As the diagonal blocks of $H_{\text{BdG}}(\bm{k})$ (Eq.~\eqref{eq_supp: H_bdg}) do not depend on the order parameters, we also get

\begin{align}
    \partial_{\Delta_\mu^R} H_{\text{BdG}}(\bm{k}) &= \begin{pmatrix}
        0 & \delta \Delta_\mu (\bm{k})\\
        (\delta \Delta_\mu (\bm{k}))^\dagger & 0
    \end{pmatrix},
    \label{eq_supp: d_delta_H_R}\\
    \partial_{\Delta_\mu^I} H_{\text{BdG}}(\bm{k}) &= \begin{pmatrix}
        0 & i \delta \Delta_\mu (\bm{k})\\
        -i (\delta \Delta_\mu (\bm{k}))^\dagger & 0
    \end{pmatrix},\label{eq_supp: d_delta_H_I}
\end{align}
where we have defined $\delta \Delta_\mu (\bm{k}) \equiv \partial_{\Delta_\mu^R} \Delta_{\bm{k}}$, and $R$ and $I$ refer to real and imaginary parts, respectively. For on-site interactions we have $\Delta_{\bm{k}} = \text{diag}(\overline{\Delta}_{0 \alpha 0 \alpha})$, and thus simply $[\delta \Delta_\alpha (\bm{k})]_{\alpha \alpha} = 1$, with all of the other elements being zero, while for off-site interactions $\bm{k}$-dependent off-diagonal exponential terms arise from Eq.~\eqref{eq_supp: delta_k}. For example, for the nearest-neighbor interaction with $[\Delta_{\bm{k}}]_{\alpha \beta} = -\sum_j \overline{\Delta}_{0 \alpha j \beta} e^{i \bm{k} \cdot \bm{r}^\Delta_{0 \alpha j \beta}}$ the non-zero terms of $\delta \Delta_{0 \alpha j \beta} (\bm{k})$ are $[\delta \Delta_{0 \alpha j \beta}(\bm{k})]_{\alpha \beta} = e^{i \bm{k} \cdot \bm{r}^\Delta_{0 \alpha j \beta}}$ and $[\delta \Delta_{0 \alpha j \beta} (\bm{k})]_{\beta \alpha} = e^{i \bm{k} \cdot \bm{r}^\Delta_{j \beta 0 \alpha}} = e^{-i \bm{k} \cdot \bm{r}^\Delta_{0 \alpha j \beta}}$. The latter contribution appears due to the relation $\Delta_{0 \alpha j \beta} = \Delta_{j \beta 0 \alpha}$. Furthermore, since $\Delta_{\bm{k}}$ is linear in the order parameters, the second derivatives of $H_{\text{BdG}}(\bm{k})$ vanish.

With these ingredients, we can start to calculate quantities of the form $\braket{\phi^\pm_{\bm{k}m} | \partial_{\Delta_\mu^{R/I}} H_{\text{BdG}}(\bm{k}) | \phi^\pm_{\bm{k}n}}$. We get
\begin{align}
    \braket{\phi^+_{\bm{k}m} | \partial_{\Delta_\mu^{R/I}} H_{\text{BdG}}(\bm{k}) | \phi^+_{\bm{k}n}} &= 0\\
    \braket{\phi^-_{\bm{k}m} | \partial_{\Delta_\mu^{R/I}} H_{\text{BdG}}(\bm{k}) | \phi^-_{\bm{k}n}} &= 0\\
    \braket{\phi^+_{\bm{k}m} | \partial_{\Delta_\mu^R} H_{\text{BdG}}(\bm{k}) | \phi^-_{\bm{k}n}} &= \braket{m_{\bm{k} + \qpdw} | \delta \Delta_\mu (\bm{k}) | n_{\bm{k} - \qpdw} }\\
    \braket{\phi^+_{\bm{k}m} | \partial_{\Delta_\mu^I} H_{\text{BdG}}(\bm{k}) | \phi^-_{\bm{k}n}} &= i \braket{m_{\bm{k} + \qpdw} | \delta \Delta_\mu (\bm{k}) | n_{\bm{k} - \qpdw} }.\label{eq_supp: d_delta_H_I_matrix_element}
\end{align}
Inserting these results into Eqs. \eqref{eq_supp: dx_E}-\eqref{eq_supp: dxdy_E}, we get
\begin{widetext}
\begin{align}
    \partial_{\Delta_\mu^{R/I}} E^\pm_{\bm{k} m} |_{\Vec{\Delta} = 0} &= 0\\
    \partial_{\Delta_\mu^{R/I}} \partial_{\Delta_\nu^{R/I}} E_{\bm{k} m}^\pm |_{\Vec{\Delta} = 0} &= \pm \sum_n \frac{1}{\xi_{\bm{k} \pm \qpdw m} + \xi_{\bm{k} \mp \qpdw n}} \cdot 2 \Re \bigg( \exval{m_{\bm{k}\pm\bm{q}}|\delta \Delta_\mu^\pm (\bm{k})|n_{\bm{k}\mp\bm{q}}} \exval{n_{\bm{k}\mp\bm{q}}| \delta \Delta_\nu^\mp (\bm{k}) |m_{\bm{k}\pm\bm{q}}} \bigg)\\
    \partial_{\Delta_\mu^{R}} \partial_{\Delta_\nu^{I}} E_{\bm{k} m}^\pm |_{\Vec{\Delta} = 0} &= \sum_n \frac{1}{\xi_{\bm{k} \pm \qpdw m} + \xi_{\bm{k} \mp \qpdw n}} \cdot 2 \Im \bigg( \exval{m_{\bm{k}\pm\bm{q}}|\delta \Delta_\mu^\pm (\bm{k})|n_{\bm{k}\mp\bm{q}}} \exval{n_{\bm{k}\mp\bm{q}}| \delta \Delta_\nu^\mp (\bm{k}) |m_{\bm{k}\pm\bm{q}}} \bigg),
\end{align}
where $\delta \Delta_\mu^+ (\bm{k}) \equiv \delta \Delta_\mu (\bm{k})$ and $\delta \Delta_\mu^- (\bm{k}) \equiv \delta \Delta_\mu^\dagger (\bm{k})$. Then, finally  Eq.~\eqref{eq_supp: omega_derivative_general} yields
\begin{align}
    \partial_{\Delta_\mu^{R}} \partial_{\Delta_\nu^{R}} \Omega |_{\Vec{\Delta} = 0} 
    &= \sum_{\bm{k}mn} \frac{n_F(\xi_{\bm{k} + \qpdw m}) + n_F(\xi_{\bm{k} - \qpdw n}) - 1}{\xi_{\bm{k} + \qpdw m} + \xi_{\bm{k} - \qpdw n}} \Re \bigg( \exval{m_{\bm{k}+\bm{q}}|\delta \Delta_\mu(\bm{k})|n_{\bm{k}-\bm{q}}} \exval{n_{\bm{k}-\bm{q}}| \delta \Delta_\nu^\dagger (\bm{k})|m_{\bm{k}+\bm{q}}} \bigg)\\
    &+ \sum_{\bm{k}mn} \frac{1 - n_F(\xi_{\bm{k} - \qpdw m}) - n_F(\xi_{\bm{k} + \qpdw n})}{-\xi_{\bm{k} - \qpdw m} - \xi_{\bm{k} + \qpdw n}} \Re \bigg( \exval{m_{\bm{k}-\bm{q}}|\delta \Delta_\mu^\dagger (\bm{k})|n_{\bm{k}+\bm{q}}} \exval{n_{\bm{k}+\bm{q}}| \delta \Delta_\nu (\bm{k}) |m_{\bm{k}-\bm{q}}} \bigg) + 2 N_c \frac{s_\mu}{J_\mu} \delta_{\mu \nu} \nonumber\\
    &= \sum_{\bm{k}mn} \frac{n_F(\xi_{\bm{k} + \qpdw m}) + n_F(\xi_{\bm{k} - \qpdw n}) - 1}{\xi_{\bm{k} + \qpdw m} + \xi_{\bm{k} - \qpdw n}} \cdot 2 \Re \bigg( \exval{m_{\bm{k}+\bm{q}}|\delta \Delta_\mu (\bm{k})|n_{\bm{k}-\bm{q}}} \exval{n_{\bm{k}-\bm{q}}| \delta \Delta_\nu^\dagger (\bm{k})|m_{\bm{k}+\bm{q}}} \bigg) + 2 N_c \frac{s_\mu}{J_\mu} \delta_{\mu \nu},
\end{align}
\end{widetext}
where $n_F(-E) = 1 - n_F(E)$ has been used, $\delta_{\mu \nu}$ is the Kronecker delta, and $m$ and $n$ have been swapped on the second row to get the final result. The constant $s_\mu = 1$ if $\Delta_\mu$ is an on-site order parameter. For the off-site degrees of freedom, $s_\mu = 1$ if $\Delta_{i \alpha j \beta}$ and $\Delta_{j \beta i \alpha}$ are indepedent, while $s_\mu = 2$ if $\Delta_{i \alpha j \beta} = \Delta_{j \beta i \alpha}$, as then each order parameter has a double contribution in the final term of Eq.~\eqref{eq_supp: omega_derivative_general}. Similar steps can be used to show that the derivatives with respect to the imaginary parts of the order parameters have the same form, while for the mixed derivatives (with one real and one imaginary part derivative) we only need to swap the real part operator to the imaginary part operator in the first term, while the last term with the Kronecker delta vanishes by the assumption of independent real and imaginary parts of the order parameters. Thus, the final result for the Hessian matrix is
\begin{align}
    \omegahessian = \begin{pmatrix}
        \Re X & \Im X\\
        -\Im X & \Re X
    \end{pmatrix} + 2 N_c \bm{1}_{2 \times 2} \otimes \text{ diag}(s_\mu/J_\mu),
    \label{eq_supp: omega_hessian_final_result}
\end{align}
where
\begin{align}
    X_{\mu \nu} = &\sum_{\bm{k}mn} \frac{n_F(\xi_{\bm{k} + \qpdw m}) + n_F(\xi_{\bm{k} - \qpdw n}) - 1}{\xi_{\bm{k} + \qpdw m} + \xi_{\bm{k} - \qpdw n}} \nonumber \\ &\times 2 \exval{m_{\bm{k}+\bm{q}}|\delta \Delta_\mu(\bm{k})|n_{\bm{k}-\bm{q}}} \exval{n_{\bm{k}-\bm{q}}| \delta \Delta_\nu^\dagger (\bm{k})|m_{\bm{k}+\bm{q}}},
    \label{eq_supp: X}
\end{align}
and $\bm{1}_{2 \times 2}$ is the $2 \times 2$ identity matrix. In the lower left block, we have used $\Im X^T = -\Im X$, resulting from the hermiticity of $X$. When both the interaction strength $J_{\mu} = J$ and $s_\mu = s$ are taken to be constants, the second term becomes proportional to the $2 n \times 2n$ identity matrix $\bm{1}_{2n \times 2n}$, where $n$ is the number of independent order parameters. Since this contribution simply shifts the eigenvalues of the Hessian matrix, we can absorb it into the first term of Eq.~\eqref{eq_supp: omega_hessian_final_result} by writing
\begin{align}
    \omegahessian &= \frac{2 s N_c}{J} (-\chi + \bm{1}_{2n \times 2n}),\\
    \chi &= -\frac{J}{2 s N_c} \begin{pmatrix}
        \Re X & \Im X\\
        -\Im X & \Re X
    \end{pmatrix},
\end{align}
which defines the pairing susceptibility $\chi$. Each eigenvalue $\epsilon$ of $\chi$ now maps to an eigenvalue $\frac{2 s N_c}{J}(-\epsilon + 1)$ of the full Hessian. Thus, the smallest eigenvalue of the full Hessian being zero is equivalent to the largest eigenvalue of $\chi$ being one. Furthermore, it turns out that it suffices to use the matrix $-J/(2s N_c) X$ instead of the larger matrix $\chi$. In particular, for each eigenvector $v$ of $-J/(2s N_c) X$ with eigenvalue $\varepsilon$, $\chi$ has a pair of eigenvectors $(\Re v, -\Im v)^T$ and $(\Im v, \Re v)^T$ with the same eigenvalue $\varepsilon$. This redundancy reflects the global phase freedom of the order parameters. In case the ratio $s_\mu / J_\mu$ is not constant, one should use the full Hessian instead of $\chi$.

This expression for the pairing susceptibility is prone to finite-size effects at low temperatures due to the Fermi surface divergence of the DOS term in the limit $T \rightarrow 0$. Thus, to obtain accurate results, careful finite-size analysis should be carried out in numerical calculations with finite systems, especially regarding the distribution of $\bm{k}$-points near the Fermi surface.

\section{Gap Equation}

For a given $\qpdw$, the order parameters $\Delta_\mu$ are determined by minimizing the grand potential $\Omega$ \eqref{eq_supp: omega}. Thus, we obtain the condition $\partial_{\Delta_\mu^R} \Omega = \partial_{\Delta_\mu^I} \Omega = 0$. By using the Hellmann-Feynman theorem and a calculation similar to what was done in Apps. \ref{sec: supp_grand_potential} and \ref{sec: supp_pairing_susceptibility}, we can write this condition as

\begin{align}
    \Delta_\mu^R &= -\frac{J_\mu}{2 N_c s_\mu} \sum_{\bm{k} a} n_F(E_{\bm{k}a}) \braket{\psi_{\bm{k}a}| \partial_{\Delta_\mu^R} H_{\text{BdG}}(\bm{k}) | \psi_{\bm{k}a}},\\
    \Delta_\mu^I &= -\frac{J_\mu}{2 N_c s_\mu} \sum_{\bm{k} a} n_F(E_{\bm{k}a}) \braket{\psi_{\bm{k}a}| \partial_{\Delta_\mu^I} H_{\text{BdG}}(\bm{k}) | \psi_{\bm{k}a}},
\end{align}
with $s_\mu \in \{ 1, 2\}$ as defined in App. \ref{sec: supp_pairing_susceptibility}. Here, the eigenvectors of $H_{\text{BdG}}(\bm{k})$ are denoted by $\psi$ instead of $\phi$ to avoid confusion in the following. Namely, we can obtain band-resolved contributions to the order parameters by expanding the $\ket{\psi_{\bm{k}a}}$ in terms of the non-interacting eigenvectors $\ket{\phi^+_{\bm{k}m}} = (\ket{m_{\bm{k}+\qpdw}}, 0)^T$ and $\ket{\phi^-_{\bm{k}m}} = (0, \ket{m_{\bm{k}-\qpdw}})^T$ of Eqs. \eqref{eq_supp: phi_plus} and \eqref{eq_supp: phi_minus} to get

\begin{align}
    \Delta_\mu^R &= -\frac{J_\mu}{N_c s_\mu} \sum_{\bm{k}mn} \Re\bigg( c_{\bm{k} mn} \braket{m_{\bm{k}+\qpdw}|\delta \Delta_\mu(\bm{k}) |n_{\bm{k}-\qpdw}} \bigg),\\
    \Delta_\mu^I &= \frac{J_\mu}{N_c s_\mu} \sum_{\bm{k}mn} \Im\bigg( c_{\bm{k} mn} \braket{m_{\bm{k}+\qpdw}|\delta \Delta_\mu(\bm{k})  |n_{\bm{k}-\qpdw}} \bigg),
\end{align}
where

\begin{align}
    c_{\bm{k}mn} = \sum_a n_F(E_{\bm{k}a}) \braket{\psi_{\bm{k}a}|\phi^+_{\bm{k}m}} \braket{\phi^-_{\bm{k}n}|\psi_{\bm{k}a}}.
\end{align}
Here we have used Eqs. \eqref{eq_supp: d_delta_H_R}-\eqref{eq_supp: d_delta_H_I_matrix_element}. Combining these results for the real and imaginary parts yields

\begin{align}
    \Delta_\mu = -\frac{J_\mu}{N_c s_\mu} \sum_{\bm{k}mn} \big(c_{\bm{k} mn} \braket{m_{\bm{k}+\qpdw}|\delta \Delta_\mu(\bm{k})  |n_{\bm{k}-\qpdw}}\big)^*.\label{eq_supp: gap_eq_band_resolved}
\end{align}
This form allows separating the pairing contribution for each pair of bands $m, n$ and momentum $\bm{k}$. From here one can also explicitly see how intra-flat-band pairing vanishes for bipartite lattices with unequal sublattice sizes, as then $\braket{m_{\bm{k}+\qpdw}|\delta \Delta_\mu |n_{\bm{k}-\qpdw}} = 0$ with $m,n$ pointing to a flat band as discussed in the main text.

\section{Superfluid Weight from the Grand Potential}

\label{sec: supp_sfw}

We define the superfluid weight $D_s$ in terms of the grand potential as
\begin{align}
    [D_s]_{ij} = \frac{1}{V} \frac{\text{d}^2 \Omega(\bm{q}')}{\text{d} q_i' \text{d} q_j'} \bigg |_{\bm{q}' = \qpdw},
    \label{eq_supp: sfw}
\end{align}
where $V$ is the volume of the system (area in 2D). The superfluid weight can be calculated with a straightforward application of the chain rule to the function $\Omega = \Omega(\bm{q}, \Delta_\mu, \mu)$, as was done in Ref.~\cite{Huhtinen2022FlatBandSCQuantumMetric}. It is critical to note that the derivative in Eq.~\eqref{eq_supp: sfw} is the total derivative, and thus the implicit dependence of the order parameters on $\bm{q}$ must be taken into account. On the other hand, we consider our system in the grand canonical ensemble where the chemical potential is fixed (instead of the particle number being fixed) and thus ignore the dependence on $\mu$. Using the chain rule and the minimizing condition $\partial_{\Delta_\mu^R} \Omega = \partial_{\Delta_\mu^I} \Omega = 0$, we end up with
\begin{widetext}
\begin{align}
    V [D_s]_{ij} &= \frac{\partial^2 \Omega(\bm{q}')}{\partial q_i' \partial q_j'} \bigg |_{\bm{q}' = \qpdw} - \bm{f_i}^T(\bm{q}') \partial^2_\Delta \tilde{\Omega} \bm{f_j}(\bm{q}') |_{\bm{q}' = \qpdw}\\
    &= \frac{\partial^2 \Omega(\bm{q}')}{\partial q_i' \partial q_j'} \bigg |_{\bm{q}' = \qpdw} - \bm{b_i}^T(\bm{q}') [\partial^2_\Delta \tilde{\Omega}]^{-1} \bm{b_j}(\bm{q}') |_{\bm{q}' = \qpdw},\\
    \partial^2_\Delta \tilde{\Omega} &= \begin{pmatrix}
        \frac{\partial^2 \Omega}{\partial \Delta_1^R \partial \Delta_1^R} & \cdots & \frac{\partial^2 \Omega}{\partial \Delta_1^R \partial \Delta_{n}^R} & \frac{\partial^2 \Omega}{\partial \Delta_1^R \partial \Delta_2^I} & \cdots & \frac{\partial^2 \Omega}{\partial \Delta_1^R \partial \Delta_{n}^I}\\
        \hdots & \ddots & \hdots & \hdots & \ddots & \hdots\\
        \frac{\partial^2 \Omega}{\partial \Delta_{n}^R \partial \Delta_1^R} & \cdots & \frac{\partial^2 \Omega}{\partial \Delta_{n}^R \partial \Delta_{n}^R} & \frac{\partial^2 \Omega}{\partial \Delta_{n}^R \partial \Delta_2^I} & \cdots & \frac{\partial^2 \Omega}{\partial \Delta_{n}^R \partial \Delta_{n}^I}\\
        \frac{\partial^2}{\partial \Delta_2^I \partial \Delta_1^R} & \cdots & \frac{\partial^2}{\partial \Delta_2^I \partial \Delta_{n}^R} & \frac{\partial^2}{\partial \Delta_2^I \partial \Delta_2^I} & \cdots & \frac{\partial^2}{\partial \Delta_2^I \partial \Delta_{n}^I}\\
        \hdots & \ddots & \hdots & \hdots & \ddots & \hdots\\
        \frac{\partial^2}{\partial \Delta_{n}^I \partial \Delta_1^R} & \cdots & \frac{\partial^2}{\partial \Delta_{n}^I \partial \Delta_{n}^R} & \frac{\partial^2}{\partial \Delta_{n}^I \partial \Delta_2^I} & \cdots & \frac{\partial^2}{\partial \Delta_{n}^I \partial \Delta_{n}^I}\\
    \end{pmatrix},\\
    \bm{f_i}(\bm{q}') &= \begin{pmatrix} \frac{\intd{\Delta_1^R}}{\intd{q_i'}} &
    \cdots & \frac{\intd{\Delta_n^R}}{\intd{q_i'}} &\frac{\intd{\Delta_2^I}}{\intd{q_i'}}
    &\cdots
    &\frac{\intd{\Delta_n^I}}{\intd{q_i'}}
    \end{pmatrix}^T,
    \label{eq_supp: delta_q_derivatives_definition}\\
    \bm{b_i}(\bm{q}') &= \begin{pmatrix} \frac{\partial^2 \Omega}{\partial q_i' \partial \Delta_1^R} &
    \cdots & \frac{\partial^2 \Omega}{\partial q_i' \partial \Delta_{n}^R} &\frac{\partial^2 \Omega}{\partial q_i' \partial \Delta_2^I}
    &\cdots
    &\frac{\partial^2 \Omega}{\partial q_i' \partial \Delta_{n}^I}
    \end{pmatrix}^T.
\end{align}
\end{widetext}
This result is essentially identical to the one derived in Ref.~\cite{Huhtinen2022FlatBandSCQuantumMetric} for the Hubbard-$U$ interaction; the only difference is that for us, the degrees of freedom indexing the order parameters need not be just the orbitals. To obtain the second line from the first, we have used the useful relation

\begin{align}
    \partial^2_\Delta \tilde{\Omega} \bm{f_i}(\bm{q}') = - \bm{b_i}(\bm{q}')
    \label{eq_supp: delta_q_derivatives}
\end{align}
that allows calculating the derivatives of the order parameters purely in terms of properties of the ground state. Notice that the derivatives with respect to $\Delta_1^I$ are missing in $\partial^2_\Delta \tilde{\Omega}, \bm{b_i}$, and $\bm{f_i}$. This is because we take $\Delta_1$ to be real to fix the overall phase; otherwise $\partial^2_\Delta \tilde{\Omega}$ would not be invertible. It should also be noted that unlike the similar expression used when calculating the pairing susceptibility, $\partial^2_\Delta \tilde{\Omega}$ is not evaluated at $\Vec{\Delta} = 0$, but instead at the ground-state values of the order parameters. This expression is numerically more convenient than Eq.~\eqref{eq_supp: sfw} since it only contains partial derivatives. Evaluating the total derivative with e.g. finite-difference methods would require solving the order parameters for many different values of $\bm{q}$, but with the partial derivatives, this is not necessary. However, we can do even better and calculate all of the necessary derivatives using the Hellmann-Feynman theorem and general result in Eq.~\eqref{eq_supp: omega_derivative_general}, which avoids numerical differentiation altogether by shifting all of the derivatives to the BdG Hamiltonian. Such derivatives are simple to take analytically; in particular, the $\bm{q}$-derivatives have the following form (for any interaction)
\begin{align}
    \partial_{q_i} H_{\text{BdG}}(\bm{k}) &= \begin{pmatrix}
        \partial_{k_i} H_{\bm{k}+\qpdw} & 0\\
        0 & \partial_{k_i} H_{\bm{k}-\qpdw}
    \end{pmatrix},\\
    \partial_{q_i} \partial_{q_j} H_{\text{BdG}}(\bm{k}) &= \begin{pmatrix}
        \partial_{k_i} \partial_{k_j} H_{\bm{k}+\qpdw} & 0\\
        0 & -\partial_{k_i} \partial_{k_j} H_{\bm{k}-\qpdw}
    \end{pmatrix},\\
    \partial_{k_i} H_{\bm{k}\pm\qpdw} &= \frac{\partial H_{\bm{k'}}}{\partial k_i'} \bigg |_{\bm{k'} = \bm{k} \pm \qpdw},\\
    \partial_{k_i} \partial_{k_j} H_{\bm{k}\pm\qpdw} &= \frac{\partial^2 H_{\bm{k'}}}{\partial k_i' \partial k_j'} \bigg |_{\bm{k'} = \bm{k} \pm \qpdw},
\end{align}
while the derivatives with respect to the order parameters are given in Eqs. \eqref{eq_supp: d_delta_H_R}-\eqref{eq_supp: d_delta_H_I}. These results can then be plugged into Eq.~\eqref{eq_supp: omega_derivative_general} to obtain the derivatives of the grand potential that yield the superfluid weight.

{ 

\section{Superfluid Weight from Linear Response Theory}

\label{sec: supp_sfw_lin_response}

\renewcommand{\mu}{x}
\renewcommand{\nu}{y}

In this section, we derive the superfluid weight from linear response theory for the mean-field Hubbard model with nearest-neighbor pairing. The superfluid weight is here defined via the linear response to an external vector potential $\bm{A}$ that is introduced using a Peierls substitution in the kinetic term of the Hamiltonian by modifying the hopping amplitudes as $t^\sigma _{i\alpha, j\beta}(\bm{A}) = t^\sigma _{i\alpha, j\beta} \exp({-i \int_{\bm{r}_{j\beta} }^{\bm{r}_{i \alpha}} \bm{A}(\bm{r},t) \cdot d\bm{r}}).$ We assume that $\bm{A} (\bm{r} ,t) = \bm{A}_{\mathrm{amp}} \exp(i \tilde{\bm{q}} \cdot \bm{r} -i \omega t)$ varies slowly in both space and time, giving the response current the same spatial and temporal form, given as $\mathbf{j} (\bm{r} ,t) = \mathbf{j}_{\mathrm{amp}} \exp(i \tilde{\bm{q}} \cdot \bm{r} -i \omega t)$. From this it also follows that the hopping term can be approximated by $t^\sigma _{i\alpha j\beta}(\bm{A}) = t^\sigma _{i\alpha, j\beta} e^{i \bm{A} (\bm{r}^{CM}_{i\alpha, j\beta} ,t) \cdot \bm{r}^{\Delta}_{i\alpha, j\beta}},$ where $\bm{r}^{CM}_{i\alpha, j\beta} = (\bm{r}_{i\alpha} + \bm{r}_{j\beta})/2$ is the center-of-mass position of the two sites and $\bm{r}^{\Delta}_{i\alpha, j\beta} = \bm{r}_{j\beta} - \bm{r}_{i\alpha}$ is the corresponding relative position.

It should be noted that the wave vector $\tilde{\bm{q}}$, describing the spatial variation of $\bm{A}$, is distinct from the PDW wave vector $\qpdw$. Eventually, we are interested in the static long-wavelength limit $\omega, \tilde{\bm{q}} \rightarrow 0$, in which case $\bm{A}$ becomes a constant. The Peierls substitution is then equivalent to substituting $H_{\bm{k}}^\sigma \rightarrow H_{\bm{k} + \bm{A}}^\sigma$ such that $\bm{A}$ itself becomes equivalent to $\bm{q}$, which relates the linear response definition of superfluid weight to the one given in terms of grand potential (namely, the two definitions are equivalent)~\cite{Huhtinen2022FlatBandSCQuantumMetric}. We will comment on the equivalence again at the end of this section, but for now, we continue with a general $\bm{A}$ with the time and space dependencies included.


The total current density induced by the vector potential is given by $j_\mu (\bm{r},t) = -\delta H (\mathbf{A})/\delta A_\mu (\bm{r},t)$, where $\delta/ \delta A_{\mu}$ denotes a functional derivative. To avoid overloading notation, in this section, we denote generic spatial coordinates as $x$ and $y$, with the understanding that they could be replaced with any combination of $x,y,z$, including $xx$, etc. Now, the kinetic Hamiltonian has a contribution to the current density given by
\begin{widetext}
\begin{align}
    \frac{\delta H_{\mathrm{kin}}(\boldsymbol{A})}{\delta A_{\mu}(\boldsymbol{r},t)} = \sum_{i \alpha, j \beta} \bigg( T_{\mu\nu}(i\alpha,j\beta)A_{\nu}(\boldsymbol{r},t)+j_{\mu}^{\mathrm{p}}(i\alpha,j\beta) \bigg) \delta_{\bm{r}, \bm{r}^{\text{CM}}_{i \alpha, j \beta}},
    \label{eq:H_kin func derivative}
\end{align}
where $j_\mu ^p (i\alpha, j\beta) = i \sum_\sigma t^\sigma _{i\alpha, j\beta} [r^{\Delta}_{i\alpha, j\beta}]_\mu c^{\dagger} _{i\alpha\sigma} c_{j\beta\sigma}$ and $T_{\mu \nu} (i\alpha, j\beta)  A_\nu (\bm{r}^{CM}_{i\alpha, j\beta} ,t) = - \sum_\sigma t^\sigma _{i\alpha, j\beta} [r^{\Delta}_{i\alpha, j\beta}]_\mu [r^{\Delta}_{i\alpha, j\beta}]_\nu c^{\dagger} _{i\alpha\sigma} c_{j\beta\sigma} A_\nu (\bm{r}^{CM}_{i\alpha, j\beta} ,t)$ are the paramagnetic and the diamagnetic current operators, respectively. Repeated indices are summed over.


After the mean-field approximation, the nearest-neighbor interaction term described in the main text reads

\begin{align}
    {H}_{\mathrm{int}} &= \frac{1}{2} \sum_{ i\alpha,{j}\beta} I_{i \alpha j \beta}
    \left(\Delta_{{i}\alpha {j}\beta}{h}_{{i}\alpha {j}\beta}^{\dagger}+\Delta_{{i}\alpha {j}\beta}^{*}{h}_{{i}\alpha {j}\beta}+\frac{2}{J}|\Delta_{{i}\alpha {j}\beta}|^{2}\right).
\end{align}
Here, $I_{i\alpha j\beta}$ is an indicator variable with value $1$ if $i \alpha$ and $j \beta$ are nearest neighbors and zero otherwise. The functional derivatives of the interaction term are
\begin{align}
    \frac{\delta H_{\mathrm{int}}(\boldsymbol{A})}{\delta A_{\mu}(\boldsymbol{r},t)} = \frac{1}{2} \sum_{{i}\alpha,{j}\beta } I_{i \alpha j \beta} \left(\frac{\delta \Delta_{i\alpha j\beta}}{\delta A_\mu}h_{i\alpha j\beta}^\dagger + \frac{\delta \Delta_{i\alpha j\beta}^*}{\delta A_\mu}h_{i\alpha j\beta} + \frac{1}{J} \frac{\delta \Delta_{i\alpha j\beta}}{\delta A_\mu} \Delta_{i\alpha j\beta}^* + \mathrm{H.c.} \right) \delta_{\bm{r}, \bm{r}^{\text{CM}}_{i \alpha, j \beta}}.  
    \label{eq:H_int func derivative}
\end{align}
Using a linear response approximation, we can express the order parameter as $\Delta_{i\alpha j\beta}(\mathbf{A}) \approx \Delta_{i\alpha j\beta}(\mathbf{A=0}) + \delta \Delta_{i\alpha j\beta}/\delta A_\nu |_{\mathbf{A=0}} A_\nu$, and rewrite Eq.~\eqref{eq:H_int func derivative} as
\begin{equation}
    \begin{aligned}
        \frac{\delta H_{\mathrm{int}}(\boldsymbol{A})}{\delta A_{\mu}(\boldsymbol{r},t)} =& \frac{1}{2} \sum_{{i}\alpha,{j}\beta } I_{i \alpha j \beta} \left(\frac{\delta \Delta_{i\alpha j\beta}}{\delta A_\mu}h_{i\alpha j\beta}^\dagger + \text{H.c.} \right) \delta_{\bm{r}, \bm{r}^{\text{CM}}_{i \alpha, j \beta}}\\
        &+ \frac{1}{J}\sum_{i \alpha, j \beta} I_{i \alpha j \beta} \left(\Delta_{i\alpha j\beta}^*\frac{\delta \Delta_{i\alpha j\beta}}{\delta A_\mu} + \text{H.c.} \right) \delta_{\bm{r}, \bm{r}^{\text{CM}}_{i \alpha, j \beta}}
        + \frac{1}{J}\sum_{i \alpha, j \beta} I_{i \alpha j \beta} \left(\frac{\delta \Delta_{i\alpha j\beta}}{\delta A_\mu}\frac{\delta \Delta_{i\alpha j\beta}^*}{\delta A_\nu} + \text{H.c.}  \right) A_\nu \delta_{\bm{r}, \bm{r}^{\text{CM}}_{i \alpha, j \beta}}.
        \label{eq: H_int after lin response approx}
    \end{aligned}
\end{equation}
We use the shorthand notation $\delta\Delta_{i \alpha j \beta}/\delta A_{\mu}=\delta\Delta_{i\alpha j \beta}/\delta A_{\mu}(\boldsymbol{r}^{CM} _{i \alpha, j \beta},t)\big|_{\boldsymbol{A}=\boldsymbol{0}}$. The total current density operator is achieved by combining equations \eqref{eq:H_kin func derivative} and \eqref{eq: H_int after lin response approx}
\begin{align}
    \langle j_{\mu}(\boldsymbol{r},t)\rangle&=-\sum_{i \alpha, j \beta}\left[\left\langle\widetilde{T}_{\mu\nu}(i\alpha,j\beta)\right\rangle  A_{\nu}(\bm{r},t)+\left\langle\widetilde{j}_{\mu}^{\mathrm{p}}(i\alpha,j\beta)\right\rangle\right] \delta_{\bm{r}, \bm{r}^{\text{CM}}_{i \alpha, j \beta}},\label{eq:currentcurrent} \\
    \widetilde{T}_{\mu\nu}(i\alpha,j\beta) &= T_{\mu\nu}(i\alpha,j\beta) + I_{i \alpha j \beta} \frac{1}{J} \left(
    \frac{\delta \Delta_{i\alpha j\beta}}{\delta A_\mu}\frac{\delta \Delta_{i\alpha j\beta}^*}{\delta A_\nu} +\text{H.c.}  \right), \\
    \widetilde{j}_{\mu} ^p (i\alpha,j\beta) &= j_{\mu}^p(i\alpha,j\beta) + \frac{1}{2} I_{i \alpha j \beta} \left(\frac{\delta \Delta_{i\alpha j\beta}}{\delta A_\mu }h_{i\alpha j\beta}^\dagger + \text{H.c.}  \right)
        + I_{i \alpha j \beta}\frac{1}{J}\left(\Delta_{i\alpha j\beta}^*\frac{\delta \Delta_{i\alpha j\beta}}{\delta A_\mu} + \text{H.c.}  \right).
\end{align}
Using the expression above, we define the current-current response function by $K_{\mu\nu}(\tilde{\bm{q}}, \omega)$ such that 
$j_\mu(\tilde{\bm{q}}, \omega) = -K_{\mu\nu} (\tilde{\bm{q}}, \omega) A_{\nu} (\tilde{\bm{q}}, \omega)$. The Fourier-transformed total current density is of the form $\langle j_{\mu}(\tilde{\bm{q}},t)\rangle=(1/V)\sum_{\boldsymbol{r}}\langle j_{\mu}(\boldsymbol{r},t)\rangle e^{-i\tilde{\bm{q}}\cdot\boldsymbol{r}}$. For readability, we will use the notation $\partial_{\mu}H^\sigma|_{\boldsymbol{k}}=\partial H^\sigma_{\bk}/\partial k_{\mu}^{\prime}|_{\boldsymbol{k}^{\prime}=\boldsymbol{k}}.$ The total current density operator \eqref{eq:currentcurrent} in momentum space reads


\begin{align}
    \langle j_{\mu}(\tilde{\bm{q}},t)\rangle &=-\left\langle\widetilde{T}_{\mu\nu}\right\rangle A_{\nu}(\tilde{\bm{q}},t)-\left\langle\widetilde{j}_{\mu}^{\mathrm{p}}(\tilde{\bm{q}})\right\rangle, \\
    \begin{split}
        \widetilde{T}_{\mu \nu} &=\frac{1}{V}\sum_{\bm{k},\sigma}\sum_{\alpha\beta}[\partial_{\mu}\partial_\nu H^{\sigma}|_{\bm{k}}]_{\alpha\beta}c_{\bm{k}\alpha\sigma}^{\dagger}c_{\bm{k}\beta\sigma}
        + \frac{1}{V_c}\frac{1}{J}\sum_{j\alpha\beta}I_{0\alpha j\beta}\Bigg( \frac{\delta \overline{\Delta}_{0\alpha j\beta}}{\delta A_\mu}\frac{\delta \overline{\Delta}_{0\alpha j\beta}^*}{\delta A_\nu} + \text{H.c} \Bigg),
        \label{eq:diamagnetic mom with q}
    \end{split} \\
    \begin{split}
        \widetilde{j}_{\mu}^p(\tilde{\bm{q}}) &=\frac{1}{V}\sum_{\bm{k},\sigma}\sum_{\alpha\beta}[\partial_{\mu}H^{\sigma}|_{\bm{k}+\tilde{\bm{q}}/2}]_{\alpha\beta}c_{\bm{k}\alpha\sigma}^{\dagger}c_{\bm{k}+\tilde{\bm{q}}\beta\sigma} \\
        &+ \frac{1}{V} \sum_{\bm{k}\alpha\beta} \frac{\delta [\Delta_{-\bm{k}+{\tilde{\bm{q}}}/{2}}]_{\alpha\beta}}{\delta A_\mu} c_{\bm{k} + \qpdw-\tilde{\bm{q}}\beta\uparrow}^\dagger c_{-\bm{k} + \qpdw\alpha \downarrow}^\dagger + \frac{\delta [\Delta_{\bm{k}-{\tilde{\bm{q}}}/{2}}]_{\beta\alpha}^*}{\delta A_\mu} c_{-\bm{k} + \qpdw\alpha\downarrow} c_{\bm{k} + \qpdw+\tilde{\bm{q}}\beta \uparrow} \\
        &+ \frac{1}{V_c} \frac{1}{J} \sum_{j \alpha \beta} I_{0 \alpha j \beta} \big(\frac{\delta \overline{\Delta}_{0\alpha j\beta}^*}{\delta A_\mu} \overline{\Delta}_{0\alpha j\beta}(\boldsymbol{0}) + \text{H.c} \big) \delta_{\tilde{\bm{q}}, \bm{0}} .\label{eq: para mg with q}
    \end{split}
\end{align}
We have defined the volume of a unit cell as $V_c = V/N_c$. 
 
We apply linear response theory to compute the paramagnetic term using the Kubo formula 
\begin{align}
   \left\langle\widetilde{j}_{\mu}^{\mathrm{p}}(\tilde{\bm{q}},\omega)\right\rangle=-iV\sum_{\nu}\int_{0}^{\infty}\mathrm{d}te^{i\omega  t}\left\langle[\widetilde{j}_{\mu}^{\mathrm{p}}(\tilde{\bm{q}},t),\widetilde{j}_{\nu}^{\mathrm{p}}(-\tilde{\bm{q}},0)]\right\rangle A_{\nu}(\tilde{\bm{q}},\omega) ,
\end{align}
and use Matsubara formalism to compute the current-current response function $K_{\mu\nu}$. The current-current correlation function in Matsubara formalism is
\begin{equation}
    \begin{aligned}
        \Pi_{\mu\nu}(\tilde{\bm{q}},i\omega_{n})\equiv-\int_{0}^{\frac{1}{k_B T}}d\tau e^{i\omega_{n}\tau}\langle T[j_{\mu}^{p}(\tilde{\bm{q}},\tau)j_{\nu}^{p}(-\tilde{\bm{q}},0)]\rangle\equiv-\int_{0}^{\frac{1}{k_B T}}d\tau e^{i\omega_{n}\tau}\Pi_{\mu\nu}(\tilde{\bm{q}},\tau),
        \label{eq: The current-current correlation function in Matsubara formalism}
    \end{aligned}
\end{equation}
where $T$ is the imaginary time ordering operator and by our definition
\begin{align}
    \Pi_{\mu\nu}(\tilde{\bm{q}},\tau)=V^{2}\left\langle T[\widetilde{j_{\mu}^{\mathrm{p}}}(\tilde{\bm{q}},\tau)\widetilde{j_{\nu}^{\mathrm{p}}}(-\tilde{\bm{q}},0)]\right\rangle.
    \label{eq:iso pii}
\end{align}
To compute $\Pi_{\mu\nu}(\tilde{\bm{q}},\tau)$, we define the following block matrices:
\begin{align}
\widetilde{H}(\bm{k})&=\begin{pmatrix}H^\uparrow_{\bm{k} + \qpdw}&\mathbf{0}\\ \mathbf{0}& -(H^\downarrow_{-\bm{k} + \qpdw})^*\end{pmatrix},\label{1 block matrix} \\
\delta_\nu\Delta_{\bm{k}}&=\begin{pmatrix}\mathbf{0}&\frac{\delta\mathbf{\Delta}_{\bm{k}}}{\delta      
 A_\nu}\\\frac{\delta\mathbf{\Delta}^{\dagger}_{\bm{k}}}{\delta A_\nu}&\mathbf{0}\end{pmatrix},\label{ham tilde matrix} \\
G^{\alpha\beta} (\tau,\bm{k}) &=-\left(\begin{array}{c c}{{\left\langle T[c_{\bm{k} + \qpdw \alpha\uparrow}(\tau)c_{\bm{k} + \qpdw \beta\uparrow}^{\dagger}(0)]\right\rangle}}&{{\left\langle T[c_{\bm{k} + \qpdw \alpha\uparrow}(\tau)c_{-\bm{k} + \qpdw \beta\downarrow}(0)]\right\rangle}}\\ {{\left\langle T[c_{-\bm{k} + \qpdw \alpha\downarrow}^{\dagger}(\tau)c_{\bm{k} + \qpdw \beta\uparrow}^{\dagger}(0)]\right\rangle}}&{{\left\langle T[c_{-\bm{k} + \qpdw \alpha\downarrow}^{\dagger}(\tau)c_{-\bm{k} + \qpdw \beta\downarrow}(0)]\right\rangle}}\end{array}\right).\label{green nambu matrix}
\end{align}
The derivatives in $\delta_\nu\Delta_{\bm{k}}$ are taken elementwise. Note that $\qpdw$ above is the PDW wave vector, not $\tilde{\bm{q}}$. Equation \eqref{green nambu matrix} defines the Green's function of the system. Due to the quadratic form of the Hamiltonian in the BdG quasi-particle basis, the Green's function has the simple form $G(i\omega_n,\bm{k})=\sum_a|\phi_{\bk a}\rangle\langle\phi_{\bk a}|/(i\omega_n-E_{\bk a})$ in Matsubara space, which we take advantage of in the following. Analogous to the steps from Eq.~(D19) to Eq.~(D29) in Ref.~\cite{Huhtinen2022FlatBandSCQuantumMetric}, we calculate the total paramagnetic contribution 
\begin{align}
\Pi_{\mu\nu}(\tilde{\bm{q}},\tau) =-\sum_{\bm{k}}\mathrm{Tr}[G(-\tau,\bm{k})(\partial_{\mu}\widetilde{H}|_{\bm{k}+\tilde{\bm{q}}/2}\gamma^{z}+\delta_{\mu}\Delta_{\bm{k} + \tilde{\bm{q}}/2})
G(\tau,\bm{k}+\tilde{\bm{q}})(\partial_{{\nu}}\widetilde{H}|_{\bm{k}+\tilde{\bm{q}}/2}\gamma^{z}+\delta_{\nu}\Delta_{\bm{k} +\tilde{\bm{q}}/2})],
\label{eq: the total paramagnetic contribution}
\end{align}
where $\gamma^z$ denotes the Pauli matrix acting in the Nambu space. Fourier transforming Eq.~\eqref{eq: the total paramagnetic contribution} to Matsubara space yields
\begin{align}
    \Pi_{\mu\nu}(\tilde{\bm{q}},i\omega_n)&=\int_0^{\frac{1}{k_{B} T}}\mathrm{d}\tau e^{i\omega_n\tau}\Pi_{\mu\nu}(\tilde{\bm{q}},\tau) \\
    \begin{split}
        = &- k_{B} T \sum_{\bm{k}} \sum_{\Omega_m} \mathrm{Tr}[G(i \Omega_m,\bm{k})(\partial_{\mu}\widetilde{H}|_{\bm{k}+\tilde{\bm{q}}/2}\gamma^{z}+\delta_{\mu}\Delta_{\bm{k}+\tilde{\bm{q}}/2})
        G(i \Omega_m +i\omega_n,\bm{k}+\tilde{\bm{q}})(\partial_{{\nu}}\widetilde{H}|_{\bm{k}+\tilde{\bm{q}}/2}\gamma^{z}+\delta_{\nu}\Delta_{\bm{k}+\tilde{\bm{q}}/2})],
    \end{split}    
\end{align}
where $i\Omega_m = (2n+1)\pi /(k_B T)$ and $i\omega_n = 2n\pi /(k_B T)$ are the fermionic and bosonic Matsubara frequencies, respectively. Substituting $G(i\omega_n,\bm{k})=\sum_a|\phi_{\bk a}\rangle\langle\phi_{\bk a}|/(i\omega_n-E_{\bk a})$ for the Green's function and conducting Matsubara summation via contour integration we obtain the paramagnetic part of the current-current response function 
\begin{equation}
    \begin{aligned}
        K_{\mu \nu, \mathrm{para}}(\tilde{\bm{q}}, i\omega_n) &=-\frac{1}{V}\sum_{\bm{k},a,b}\frac{n_{F}(E_{\bk a})-n_{F}(E_{\bk + \tilde{\bm{q}}b})}{E_{\bk + \tilde{\bm{q}}b}-E_{\bk a}-i\omega_{n}} \\
        & \times\langle\phi_{\bk a}|(\partial_{{\mu}}\widetilde{H}|_{\bm{k}+\tilde{\bm{q}}/2}\gamma^{z}+\delta_{\mu}\Delta_{\bm{k}+\tilde{\bm{q}}/2})|\phi_{\bk + \tilde{\bm{q}}b}\rangle\langle\phi_{\bk +\tilde{\bm{q}}b}|(\partial_{{\nu}}\widetilde{H}|_{\bm{k}+\tilde{\bm{q}}/2}\gamma^{z}+\delta_{\nu}\Delta_{\bm{k}+\tilde{\bm{q}}/2})|\phi_{\bk a}\rangle,
        \label{eq: paramagnetic part of the current-current response function}
    \end{aligned}
\end{equation}
where $\begin{matrix}n_F(E)&=&1/(e^{ E/(k_B T)}+1)\end{matrix}$ is the Fermi-Dirac distribution. 

The procedure of computing the diamagnetic part to the current is straightforward. In Matsubara space, it reads
\begin{equation}
    K_{\mu \nu, \mathrm{dia}}(i\omega_n) = -\frac{k_B T}{V} \sum_{\boldsymbol{k}} \sum_{\Omega_n} \mathrm{Tr} [\partial_\mu \widetilde{H}|_{\boldsymbol{k}}  \partial_\nu G(i \Omega_n,\boldsymbol{k})] \\
    + \frac{1}{V_c}\frac{1}{J}\sum_{j\alpha\beta}I_{0\alpha j\beta}\Bigg( \frac{\delta \Delta_{0\alpha j\beta}}{\delta A_\mu}\frac{\delta \Delta_{0\alpha j\beta}^*}{\delta A_\nu} + \text{H.c} \Bigg) .
\end{equation}
We make use of the following identity $\partial_{\mu}G=-G\partial_{\mu}G^{-1}G$ and take the Matsubara summation via contour integration to obtain
\begin{equation}
    \begin{aligned}        
        K_{\mu \nu, \mathrm{dia}}(i\omega_n) =& \frac{1}{V}\sum_{\bm{k},a,b}\frac{n_{F}(E_{\bk a})-n_{F}(E_{b}(\bk))}{E_{\bk b}-E_{\bk a} - i\omega_n} \langle\phi_{\bk a}|(\partial_{{\mu}}\widetilde{H}|_{\bm{k}})|\phi_{\bk b}\rangle\langle\phi_{\bk b}|(\partial_{{\nu}}{H}_{\mathrm{BdG}}(\bk))|\phi_{\bk a}\rangle \\ &+ \frac{1}{V_c}\frac{1}{J}\sum_{j\alpha\beta}I_{0\alpha j\beta}\Bigg( \frac{\delta \overline{\Delta}_{0\alpha j\beta}}{\delta A_\mu}\frac{\delta \overline{\Delta}_{0\alpha j\beta}^*}{\delta A_\nu} + \text{H.c} \Bigg).
        \label{eq: diamagnetic part of the current current response function}
    \end{aligned}
\end{equation}
The definition for the nearest-neighbour pairing superfluid weight is $D_{\mu\nu}^{\mathrm{NN}} = \lim\limits_{\tilde{\bm{q}}\to0}\lim\limits_{\omega\to0}K_{\mu\nu}(\tilde{\bm{q}},i\omega)$.
By combining equations~\eqref{eq: paramagnetic part of the current-current response function} and~\eqref{eq: diamagnetic part of the current current response function}, we finally obtain
\begin{subequations}\label{eq: sfw rvb lin response corrections}
\begin{align}
    D_{\mu\nu}^{\mathrm{NN}}
        =&\frac{1}{V}\sum_{\bm{k},a,b}\frac{n_{F}(E_{\bk a})-n_{F}(E_{\bk b})}{E_{\bk b}-E_{\bk a}}\Bigg[ \langle\phi_{\bk a}|(\partial_{{\mu}}\widetilde{H}|_{\bm{k}})|\phi_{\bk b}\rangle\langle\phi_{\bk b}|(\partial_{{\nu}} {H}_{\mathrm{BdG}}(\bk ))|\phi_{\bk a}\rangle \label{eq: rvb sfw 1} \\
        &- \langle\phi_{\bk a}|(\partial_{{\mu}}\widetilde{H}|_{\bm{k}}\gamma^{z}+\delta_{\mu}\Delta_{\bm{k}})|\phi_{\bk b}\rangle\langle\phi_{\bk b}|(\partial_{{\nu}}\widetilde{H}|_{\bm{k}}\gamma^{z}+\delta_{\nu}\Delta_{\bm{k}})|\phi_{\bk a}\rangle \Bigg] \label{eq: rvb sfw 2}
        \\
         &+ \frac{1}{V_c}\frac{1}{J}\sum_{j\alpha\beta}I_{0\alpha j\beta}\Bigg( \frac{\delta \overline{\Delta}_{0\alpha j\beta}}{\delta A_\mu}\frac{\delta \overline{\Delta}_{0\alpha j\beta}^*}{\delta A_\nu} + \text{H.c} \Bigg),\label{eq: rvb sfw 3}
\end{align}
\end{subequations}
\end{widetext}
where the prefactor should be understood as $-\partial_E n_F(E)$ if $E_a = E_b$. As mentioned at the beginning of this section, in the limit $\omega, \tilde{\bm{q}} \rightarrow 0$ the functional relationship of the Hamiltonian with respect to the constant vector potential $\bm{A}$ becomes equivalent to that of the PDW (or order parameter phase) wave vector $\qpdw$, that is $H(\bm{A}) = H(\qpdw)$. Thus, we can make the replacement

\begin{align}
    \left.\frac{\delta \overline{\Delta}_{0\alpha j\beta}}{\delta A_\mu} \right \vert_{\bm{A} = 0} = \left. \frac{\intd{\overline{\Delta}_{0\alpha j\beta}}} {\intd{q'_\mu}} \right \vert_{\bm{q}' = \qpdw}
\end{align}
both in Eq.~\eqref{eq: rvb sfw 3} and inside the matrices $\delta_{\mu}\Delta_{\bm{k}}$. These derivatives are then the same as in Eq.~\eqref{eq_supp: delta_q_derivatives_definition}, and can be calculated using only ground-state properties via Eq.~\eqref{eq_supp: delta_q_derivatives}. Notice that the derivative on the right-hand side should be evaluated at the ground state, i.e. at $\bm{q}' = \qpdw$.


\section{Conventional and Geometric Contributions of the Superfluid Weight}

\label{sec: supp_sfw_conv_geom}


The superfluid weight is frequently classified in terms of the geometric properties of the normal state and split into so-called geometric and conventional parts~\cite{liangBandGeometryBerry2017, peottaSuperfluidityTopologicallyNontrivial2015}. Compared to a model with only on-site interactions, the $\boldsymbol{k}$ dependence of the order parameter matrix $\Delta_{\bm{k}}$ arising from nearest-neighbor interactions will result in additional terms, which is evident by decomposing the BdG Hamiltonian matrix term in Eq.~\eqref{eq: rvb sfw 1} as
\begin{align}
    &\langle\phi_{\bk a}|(\partial_{\mu}\widetilde{H}|_{\boldsymbol{k}})|\phi_{\bk b}\rangle\langle\phi_{\bk b}|(\partial_{\nu}H_{\mathrm{BdG}}|_{\boldsymbol{k}})|\phi_{\bk a}\rangle \\ &= \langle\phi_{\bk a}|(\partial_{\mu}\widetilde{H}|_{\boldsymbol{k}})|\phi_{\bk b}\rangle\langle\phi_{\bk b}|(\partial_{\nu}\widetilde{H}|_{\boldsymbol{k}})|\phi_{\bk a}\rangle  \\
    &+ \langle\phi_{\bk a}|(\partial_{\mu}\widetilde{H}|_{\boldsymbol{k}})|\phi_{\bk b}\rangle\langle\phi_{\bk b}|\partial_{\nu} \left(\begin{array}{cc}{0}&{\boldsymbol{\Delta_{k}}}\\{\boldsymbol{\Delta_{k}}^{\dagger}}&{0}\\\end{array}\right)|\phi_{\bk a}\rangle .
    \label{bdg hajotelma}
\end{align}
We will at first ignore these additional terms while classifying the rest of the superfluid weight to its conventional and geometric parts. For simplicity, here we focus only on the non-PDW case with $\bm{q} = 0$, but the generalization to PDWs with non-zero $\bm{q}$ is straight-forward. Analogous to Ref.~\cite{Huhtinen2022FlatBandSCQuantumMetric}, we expand the eigenvectors of $H_{\mathrm{BdG}}$ in terms of the Bloch functions: $|\phi_{\bk a}\rangle=\sum_{n=1}^{n_{\mathrm{orb}}}(w_{+,an \bk}|+\rangle\otimes|n_{\boldsymbol{k}\uparrow}\rangle+w_{-,an \bk}|-\rangle\otimes|n_{-\boldsymbol{k}\downarrow}^*\rangle)$. Using this relationship, we first define the usual conventional term independent of $\Delta_{\boldsymbol{k}}$
\begin{widetext}
\begin{align}
    [D^{\mathrm{conv}}]_{\mu \nu } &=  2\sum_{\bk} \sum_{nm} C_{mm}^{nn}[j_{\mu}^\uparrow(\boldsymbol{k})]_{nn}[j_{\nu}^\downarrow(-\boldsymbol{k})]_{mm} +(\mu\leftrightarrow\nu) \\
    &= 2\sum_{\bk} \sum_{ nm} C_{mm}^{nn} \partial_{{\mu}}\epsilon_{\boldsymbol{k},n,\uparrow}\partial_{{\nu}}\epsilon_{-\boldsymbol{k},m,\downarrow} +(\mu\leftrightarrow\nu),
\end{align}
where
\begin{align}
C_{ls}^{nm}&=\sum_{ab}\frac{n_{F}(E_{\bk a})-n_{F}(E_{\bk b})}{E_{\bk b}-E_{\bk a}}w_{+,an\bk}^{*}w_{+,bm\boldsymbol{k}}w_{-,bl\boldsymbol{k}}^{*}w_{-,as\boldsymbol{k}},\\
[j_{\mu}^{\sigma}(\boldsymbol{k})]_{nm}&=\langle n_{\boldsymbol{k}\sigma}|\partial_{k_{\mu}}H_{\boldsymbol{k}}^{\sigma}|m_{\boldsymbol{k}\sigma}\rangle\\
&=\delta_{nm}\partial_{{\mu}}\epsilon_{\boldsymbol{k},n,\sigma}+(\epsilon_{\boldsymbol{k},n,\sigma}-\epsilon_{\boldsymbol{k},m,\sigma})\langle\partial_{{\mu}}n_{\boldsymbol{k}\sigma}|m_{\boldsymbol{k}\sigma}\rangle.
\end{align}
This result is the same as in previous literature~\cite{rossi2021quantum}. The usual conventional part contains only diagonal elements of the current operator $j_{\mu}^{\sigma}$. It is the only term present in single-band models with on-site pairing and vanishes in the flat-band limit. However, the additional terms resulting from Eq.~\eqref{bdg hajotelma} include both conventional and geometric contributions, and thus we will redefine the conventional part later.

By our classification, the geometric contribution to the superfluid weight is a fully multiband component that can be non-zero on flat bands. Excluding the terms due to the $\bk$ dependence of $\Delta_{\bk}$, we split the remaining components into three terms:
\begin{align}
    [D^{\mathrm{geom,1}}]_{\mu \nu } =&  2\sum_{\boldsymbol{k}}  \sum_{n\neq m\atop l\neq s}C_{ls}^{nm}[j_{\mu}^{\uparrow}(\boldsymbol{k})]_{nm}[j_{\nu}^{\downarrow}(-\boldsymbol{k})]_{sl} +(\mu\leftrightarrow\nu), \\    
    [D^{\mathrm{geom,2}}]_{\mu \nu } =& 2\sum_{\boldsymbol{k}}\sum_{n \atop l\neq s }C_{ls}^{nn}[j_{\mu}^{\uparrow}(\boldsymbol{k})]_{nn}[j_{\nu}^{\downarrow}(-\boldsymbol{k})]_{sl}
    +2\sum_{\boldsymbol{k}}\sum_{n\neq m \atop l}C_{ll}^{nm}[j_{\mu}^{\uparrow}(\boldsymbol{k})]_{nm}[j_{\nu}^{\downarrow}(-\boldsymbol{k})]_{ll} +(\mu\leftrightarrow\nu), \\
    \begin{split}
    [D^{\mathrm{geom,3}}]_{\mu \nu } =& -\sum_{\boldsymbol{k},ab}\frac{n_{F}(E_{\bk a})-n_{F}(E_{\bk b})}{E_{\bk b}-E_{\bk a}}\big(\langle\phi_{\bk a}|\delta_{\mu}\Delta_{\boldsymbol{k}}|\phi_{\bk b}\rangle\langle\phi_{\bk b}|\delta_{\nu}\Delta_{\boldsymbol{k}}|\phi_{\bk a}\rangle\\    &+\langle\phi_{\bk a}|\delta_{\mu}\Delta_{\boldsymbol{k}}|\phi_{\bk b}\rangle\langle\phi_{\bk b}|\partial_{\nu}\widetilde{H}_{\boldsymbol{k}}\gamma^{z}|\phi_{\bk a}\rangle+\langle\phi_{\bk a}|\partial_{{\mu}}\widetilde{H}_{\boldsymbol{k}}\gamma^{z}|\phi_{\bk b}\rangle\langle\phi_{\bk b}|\delta_{\nu}\Delta_{\boldsymbol{k}}|\phi_{\bk a}\rangle\big)\\
    &-\frac{2}{JV_{c}}\sum_{j\alpha\beta} I_{0 \alpha j \beta} \Re \bigg(\frac{\mathrm{d}\overline{\Delta}_{0\alpha j\beta}}{\mathrm{d}q_{\mu}}\frac{\mathrm{d}\overline{\Delta}^*_{0\alpha j\beta}}{\mathrm{d}q_{\nu}}\bigg|_{\boldsymbol{q'}=\boldsymbol{q}} \bigg).
    \end{split}
    \label{vika rivi}
\end{align}
The second geometric term $[D^{\mathrm{geom,2}}]_{\mu \nu }$ includes both intra- and interband effects and vanishes when we have a purely intraband gap function. If the classification of the superfluid weight is based on the interband and intraband effects, this term is distinguished as a separate multigap term~\cite{Taisei2022SCmonolayerQgeom}. As in Ref.~\cite{julkuSuperfluidWeightBerezinskiiKosterlitzThouless2020}, we include $[D^{\mathrm{geom,2}}]_{\mu \nu }$ to the geometric contribution as it also reflects the quantum geometry of the Bloch electrons. The $\bm{q}$ derivatives of the order parameters on the last line of $[D^{\mathrm{geom,3}}]_{\mu \nu }$ appear only in multiband models and hence we include them in the geometric contribution.

Lastly, we turn our attention to the additional terms resulting from the $\bk$ dependence of order parameters (Eq. ~\eqref{bdg hajotelma}), dubbed $D^{\text{gap}}$. We get
\begin{equation}
    \begin{aligned}
        [D^{\mathrm{gap}}]_{\mu \nu } = \sum_{\bk}\sum_{nmls} &-\Bigl( C^{nm}_{ls} \langle n_{\bk \uparrow}| \partial_{\mu}\Delta_{\bk} | m^*_{-\bk \downarrow}\rangle +C^{nm}_{ls} \langle n^*_{-\bk \downarrow}| \partial_{\mu}\Delta^{\dagger}_{\bk} | m_{\bk \uparrow}\rangle \Bigr)[j_{\nu}^{\uparrow}(\boldsymbol{k})]_{ls} \\
    &+\Bigl( C^{nm}_{ls} \langle n_{\bk \uparrow}| \partial_{\mu}\Delta_{\bk} | m^*_{-\bk \downarrow}\rangle +C^{nm}_{ls} \langle n^*_{-\bk \downarrow}| \partial_{\mu}\Delta^{\dagger}_{\bk} | m_{\bk \uparrow}\rangle \Bigr)[j_{\nu}^{\downarrow}(\boldsymbol{-k})]_{sl}.
    \end{aligned}
\end{equation}
In terms of single-band and multiband properties, we further divide $[D^{\text{gap}}]_{\mu \nu }$ into two components. Firstly,
\begin{equation}
    \begin{aligned}
        [D^{\mathrm{gap, conv}}]_{\mu \nu }=\sum_{\boldsymbol{k}}\sum_{nml}-&\Big[C_{ll}^{nm}\langle n_{\bk \uparrow}| \partial_{\mu}\Delta_{\bk} | m^*_{-\bk \downarrow}\rangle
    + C_{ll}^{nm}\langle n^*_{-\bk \downarrow}| \partial_{\mu}\Delta^{\dagger}_{\bk} | m_{\bk \uparrow}\rangle\Big] [j_{\nu}^{\uparrow}(\boldsymbol{k})]_{ll} \\
    +&\Big[C_{ll}^{nm}\langle n_{\bk \uparrow}| \partial_{\mu}\Delta_{\bk} | m^*_{-\bk \downarrow}\rangle
    + C_{ll}^{nm}\langle n^*_{-\bk \downarrow}| \partial_{\mu}\Delta^{\dagger}_{\bk} | m_{\bk \uparrow}\rangle\Big] [j_{\nu}^{\downarrow}(\boldsymbol{-k})]_{ll},
    \end{aligned}
\end{equation}
which contains only diagonal components of the current operators and is thus included in the conventional contribution. The multiband term is defined as
\begin{equation}
    \begin{aligned}
        [D^{\mathrm{gap, geom}}]_{\mu \nu }=\sum_{\boldsymbol{k}}\sum_{n m l\neq s}-&\Big[C^{nm}_{ls}\langle n_{\bk \uparrow}| \partial_{\mu}\Delta_{\bk} | m^*_{-\bk \downarrow}\rangle
        + C^{nm}_{ls} \langle n^*_{-\bk \downarrow}| \partial_{\mu}\Delta^{\dagger}_{\bk} | m_{\bk \uparrow}\rangle \Big] [j_{\nu}^{\uparrow}(\boldsymbol{k})]_{ls} \\
        +&\Big[C^{nm}_{ls}\langle n_{\bk \uparrow}| \partial_{\mu}\Delta_{\bk} | m^*_{-\bk \downarrow}\rangle
        + C^{nm}_{ls} \langle n^*_{-\bk \downarrow}| \partial_{\mu}\Delta^{\dagger}_{\bk} | m_{\bk \uparrow}\rangle \Big] [j_{\nu}^{\downarrow}(\boldsymbol{-k})]_{sl},
    \end{aligned}
\end{equation}
\end{widetext}
which is included in the geometric contribution. Thus, the final classification of the parts of the superfluid weight with nearest-neighbor (NN) interactions is $D^{\mathrm{NN}} = D^{\mathrm{NN:conv}} + D^{\mathrm{NN:geom}}$, where the conventional and geometric contributions are $D^{\mathrm{NN:conv}}=D^{\mathrm{conv}}+D^{\mathrm{gap, conv}}$ and $D^{\mathrm{NN:geom}}=D^{\mathrm{geom,1}}+D^{\mathrm{geom,2}}+D^{\mathrm{geom,3}}+D^{\mathrm{gap, geom}}$, respectively.

}

\section{Additional Results for the Lieb Lattice}

\subsection{Numerical Results}

\begin{figure}
    \centering
    \includegraphics[width=\linewidth]{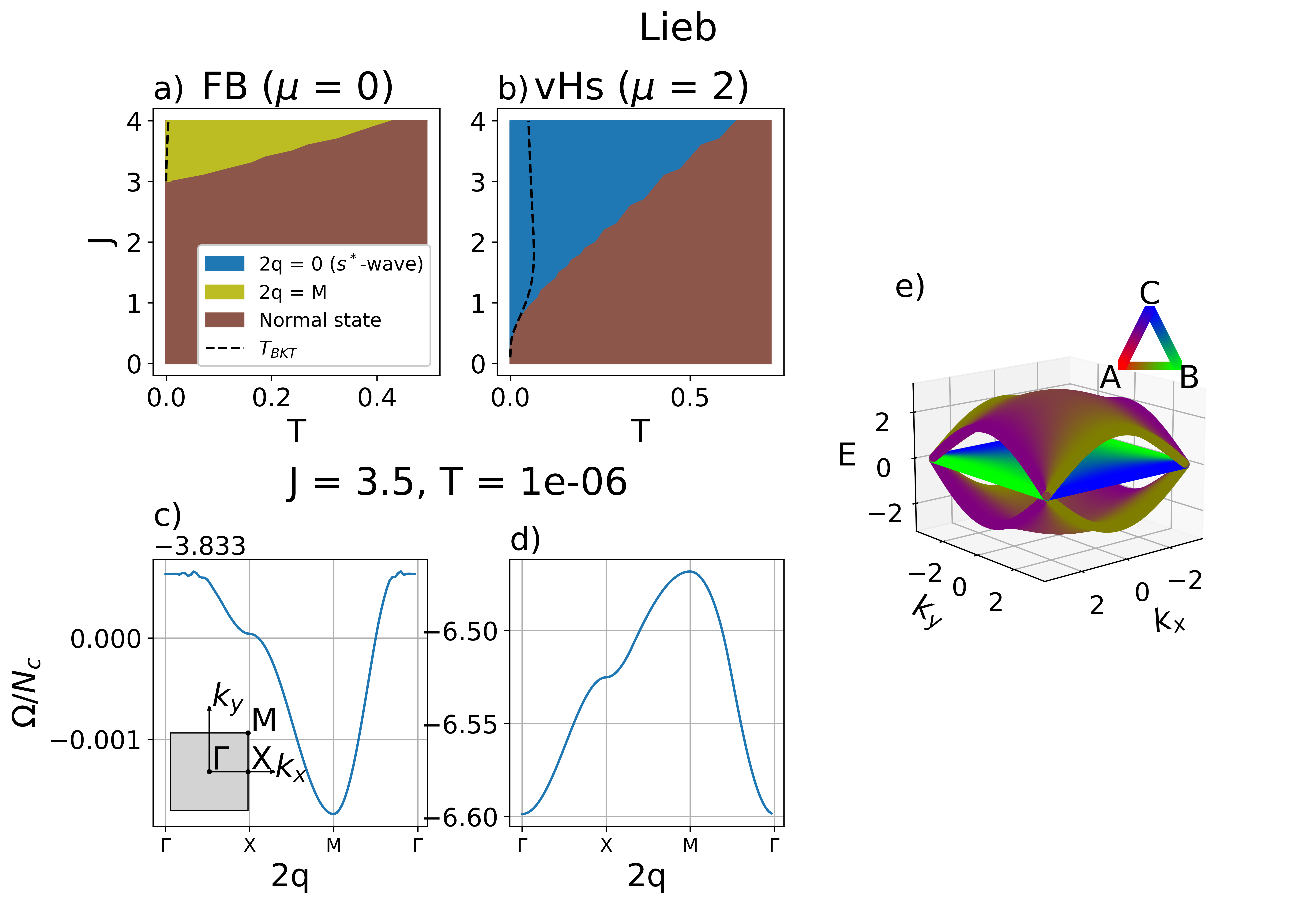}
    \caption{Phase diagrams for the Lieb lattice flat band (\textbf{a}) and vHs (\textbf{b}). The black dashed line indicates the BKT temperature. For the Lieb lattice, the $2\qpdw = \bm{M}$ and $\qpdw = 0$ are the only superconducting phases we find for the FB and vHs, respectively. As discussed in the main text, the $2\qpdw = \bm{M}$ phase preserves the $C_{4v}$ symmetry of the lattice, and the solution belongs to irrep $B_1$. On the other hand, the $\qpdw = 0$ phase at the vHs has $s^*$-wave symmetry, i.e. it is fully homogeneous in real space and belongs to irrep $A_1$. In (\textbf{a}), the finite critical interaction strength of $J_c \approx 2.9$ for the flat band is also visible. (\textbf{c}) and (\textbf{d}) show how the values of the grand potential $\Omega$ vary as a function of the PDW wave vector $2\qpdw$. (\textbf{e}) The full dispersion relation of the Lieb lattice. The colors indicate orbital composition of the Bloch states (see Fig.~\ref{fig:lieb_results} in the main text for the naming convention).}
    \label{fig:suppl_lieb_results}
\end{figure}

\begin{figure*}
    \centering
    \includegraphics[width=0.7\linewidth]{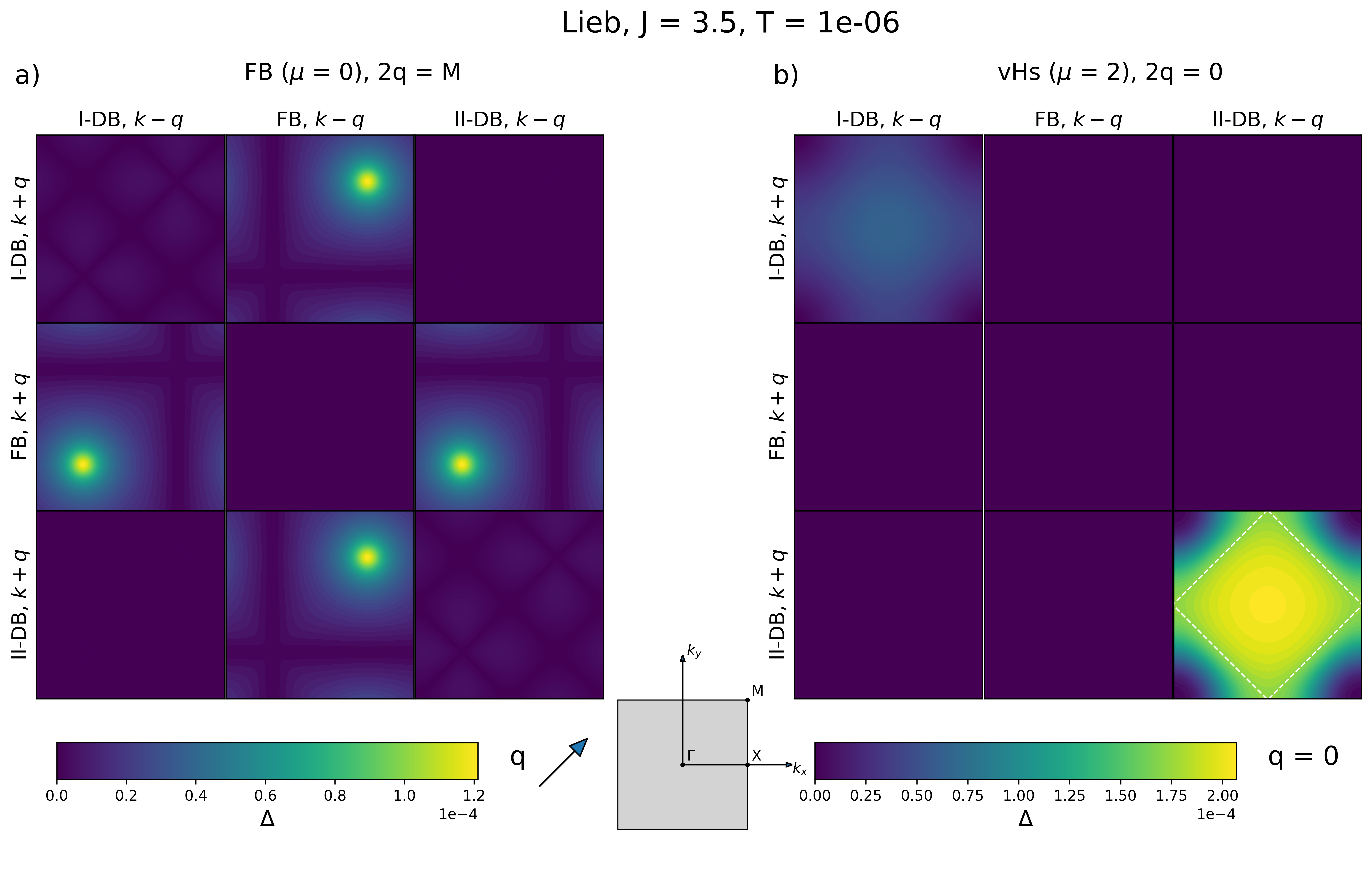}
    \caption{Band and $\bm{k}$-resolved pairing over the Brillouin zone $[-\pi, \pi]^2$ (see Eq.~\eqref{eq_supp: gap_eq_band_resolved}) for the Lieb lattice flat band $2\qpdw = \bm{M}$ PDW state (\textbf{a}), and the vHs $\qpdw = 0$ state (\textbf{b}). The blue arrow in $(\textbf{a})$ depicts $\qpdw$. The diagonal elements show intra-band pairings, and the off-diagonal the various inter-band pairings. The lower and upper dispersive bands are dubbed I-DB and II-DB, respectively. The color at $\bm{k}$ indicates the strength of pairing between $\bm{k} + \qpdw$ and $\bm{k} - \qpdw$. The white dashed line in (\textbf{b}) indicates the Fermi surface; in (\textbf{a}) there is no well-defined Fermi surface due to the flat band. Contributions from different real-space order parameters $\overline{\Delta}_{0 \alpha j \beta}$ have been combined by taking absolute values and summing over them to obtain an aggregate band and $\bm{k}$-resolved image. In (\textbf{a}), the largest contribution comes from inter-band pairing between the center of the flat band, and the Dirac points on the dispersive bands in the corners of the Brillouin zone. However, there is also a sizable contribution from other regions of the BZ due to the vanishing DOS of the Dirac points.}
    \label{fig: supp_lieb_delta_band_k_resolved}
\end{figure*}

\label{sec: supp_lieb_numerical_results}

\begin{figure}
    \centering
    \includegraphics[width=\linewidth]{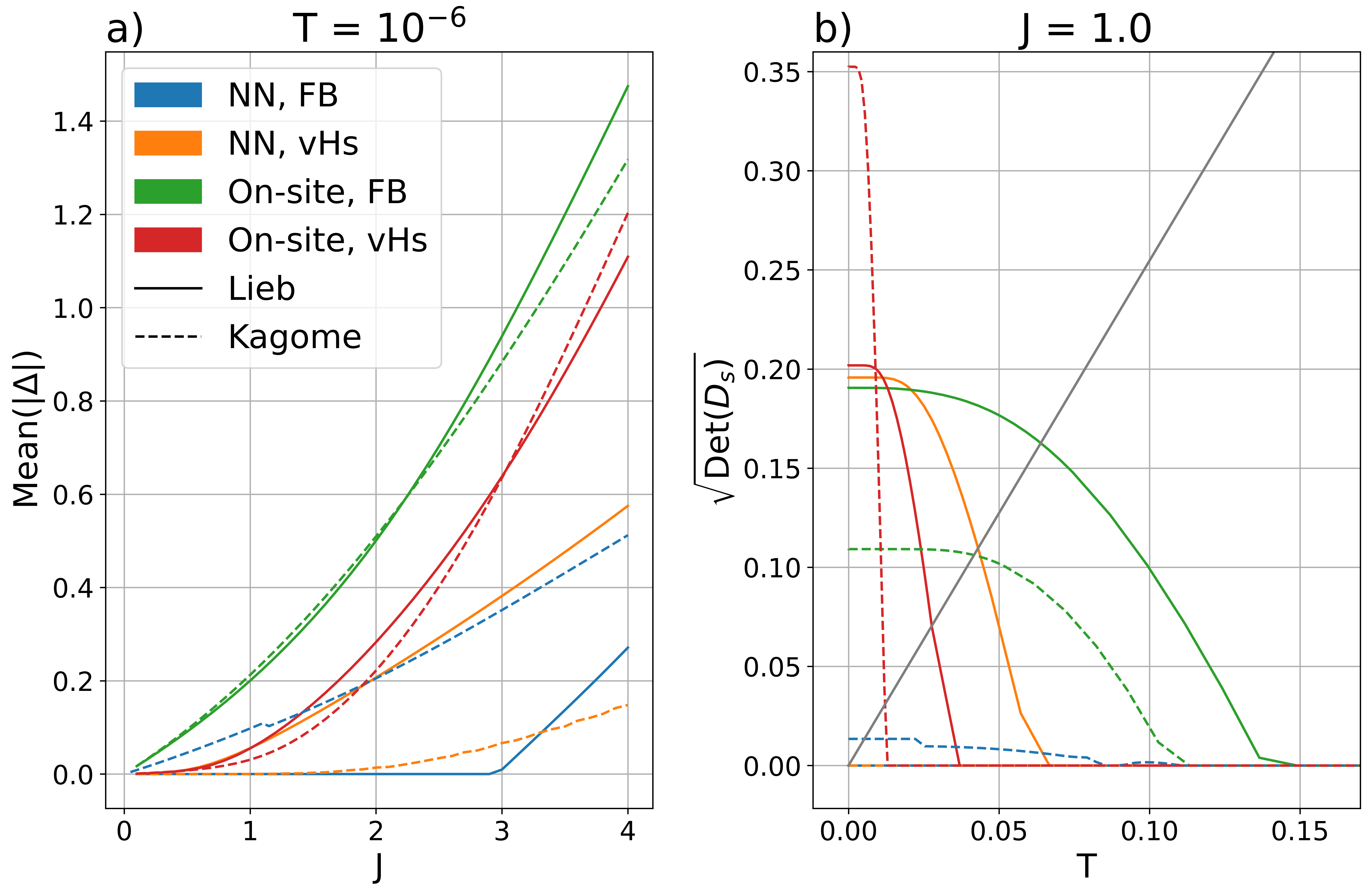}
    \caption{(\textbf{a}) The size of the order parameters in the Lieb and kagome lattice flat bands (FB) and vHs, for both on-site and nearest-neighbour (NN) interactions. Note the quantum critical point at $J \approx 2.9$ for the Lieb lattice flat band with nearest-neighbor interactions. (\textbf{b}) The temperature dependence of the superfluid weights for $J = 1$. The intersection of each graph with the gray line $y = \frac{8}{\pi} T$ indicates the corresponding BKT temperature. The BKT temperature is primarily limited by $T_c$ in the systems with more vertical intersections, while a horizontal intersection indicates limitation by the superfluid weight.}
    \label{fig:suppl_deltas_sfws}
\end{figure}

As discussed in the main text, for the Lieb lattice flat band ($\mu = 0)$ we find a $2 \qpdw = \bm{M}$ state, while the vHs ($\mu = 2$) shows a $\qpdw = 0$ state. These phases persist for a large range of temperatures and interaction strengths as seen in Figs. \ref{fig:suppl_lieb_results}(a)-(b). The values of $\qpdw$ are obtained by minimizing the grand potential, see Figs. \ref{fig:suppl_lieb_results}(c)-(d). We find that the $\qpdw = 0$ state for the vHs is mostly due to typical intra-band pairing in the upper dispersive band, while the PDW state for the flat band state consists mainly of inter-band pairing between the Dirac cones of the dispersive bands and the center of the flat band; see Fig.~\ref{fig: supp_lieb_delta_band_k_resolved}. In App. \ref{sec: supp_lieb_pdw_origin}, we provide a qualitative explanation for the PDW state. While the order parameters grow fast as function of interaction strength after the critical point $J_c \approx 2.9$ (see Fig.~\ref{fig:suppl_deltas_sfws}(a)), we find that the superfluid weight of the PDW state remains small, leading to low BKT temperatures seen in Fig.~\ref{fig: bkt_temperatures} in the main text. In contrast, the $\qpdw = 0$ state for the vHs shows order parameters and superfluid weights comparable to models with on-site interactions (see Figs. \ref{fig:suppl_deltas_sfws}(a)-(b)). An explanation for this discrepancy is explored in App. \ref{sec: supp_lieb_strong_coupling_limit}.

\subsection{Origin of the PDW State}

\label{sec: supp_lieb_pdw_origin}

In this section, we provide an approximative calculation to explain why the ground state of the Lieb lattice flat band with nearest-neighbor interactions is a PDW state with $2\qpdw = \bm{M}$. When the order parameters are small, the grand potential can be expanded as

\begin{align}
    \Omega(\Vec{\Delta}) = \Omega(\Vec{\Delta} = 0) + \frac{1}{2} \Vec{\Delta}^T \omegahessian \Vec{\Delta}.
\end{align}
Considering some small, fixed length for the vector $\Vec{\Delta}$, the grand potential is minimized when $\Vec{\Delta}$ is the eigenvector of $\omegahessian$ with the most negative eigenvalue. We assume the interaction strength $J$ to be large enough such that a negative eigenvalue exists. By the discussion at the end of App. \ref{sec: supp_pairing_susceptibility}, this is equivalent to finding the maximum eigenvalue of $-J/(2s N_c) X$, with $X$ defined in Eq.~\eqref{eq_supp: X}. The question we wish to investigate here is how $\qpdw$ affects the maximum eigenvalue.

The full expression of $X$ involves a double sum over the band indices $m,n$ and a sum (integral in the limit $N_c \rightarrow \infty$) of $\bm{k}$ over the Brillouin zone. When the temperature is small, the value of $X$ peaks heavily near the Fermi level, i.e. on the flat band. Since intra-flat-band pairing is suppressed, we consider only inter-band pairing between the flat band and the dispersive bands (see Fig.~\ref{fig: supp_lieb_delta_band_k_resolved}). For this calculation, it suffices to consider one such term with $m, n$ referring to the lower dispersive band and the flat band, respectively. Furthermore, we ignore most of the Brillouin zone and consider only a small circle of radius $k_0$ centered around the Dirac point where the dispersive band touches the Fermi level. Near the Dirac point the dispersion of the dispersive band is approximately linear, i.e. $\xi_{\bm{k}} = c |\bm{k}|$ for some $c < 0$. The contribution $S_\theta$ to the pairing susceptibility at an angle $\theta$ along this circle then becomes

\begin{align}
    S^\theta_{\mu \nu} = &-\frac{J}{4} \frac{n_F(c k_0) - \frac{1}{2}}{ck_0} \nonumber \\ &\times 2 \exval{m_\theta|\delta \Delta_\mu(\bm{k})|n_\theta^{\qpdw}} \exval{n_\theta^{\qpdw}| \delta \Delta_\nu^\dagger (\bm{k})|m_\theta},
\end{align}
where we have used $\xi_{\bm{k}} = 0$ and $n_F(0) = \frac{1}{2}$ for the flat band. Here, $\ket{m_\theta}$ is the Bloch state on the dispersive band at distance $k_0$ and angle $\theta$ from the Dirac point, while $\ket{n_\theta^{\bm{q}}}$ is the Bloch state on the flat band shifted by $-2\qpdw$ from $\ket{m_\theta}$. The dependence on $k_0$ has been suppressed, as we will focus on the dependence on $\theta$ in the following. The Bloch states can be written as

\begin{align}
    \ket{m_\theta} = \begin{pmatrix}
        \frac{1}{\sqrt{2}}\\f^m_B (\theta)\\f^m_C(\theta)
    \end{pmatrix},\qquad
    \ket{n_\theta^{\qpdw}} = \begin{pmatrix}
        0 \\ f^n_B(\theta, \qpdw)\\f^n_C(\theta, \qpdw)
    \end{pmatrix},
\end{align}
where the elements of the Bloch states are parameterized by the functions $f$, and $B,C$ refer to the orbitals (see Fig.~\ref{fig:lieb_results} in the main text). Note that the component of orbital $A$ is independent of $\bm{k}$ throughout the Brillouin zone for all bands of the Lieb lattice. Furthermore, the pairing matrix $\Delta_{\bm{k}}$ given by Eq.~\eqref{eq_supp: delta_k} is

\begin{widetext}
\begin{align}
    \Delta_{\bm{k}} = \begin{pmatrix}
        0 & \overline{\Delta}_{AB,1} g_x + \overline{\Delta}_{AB,2} g_x^* & \overline{\Delta}_{AC,1} g_y + \overline{\Delta}_{AC,2} g_y^*\\
        \overline{\Delta}_{AB,1} g_x^* + \overline{\Delta}_{AB,2} g_x & 0 & 0\\
        \overline{\Delta}_{AC,1} g_y^* + \overline{\Delta}_{AC,2} g_y & 0 & 0
    \end{pmatrix},
\end{align}
where $\overline{\Delta}_{AB, i}$ and $\overline{\Delta}_{AC, i}$ are the order parameters between the central orbital $A$ and orbitals $B$ and $C$ in the positive (for $i=1$) or negative (for $i=2$) $x$- and $y$-directions, respectively (see Fig.~\ref{fig:lieb_results}(a) in the main text), and we denote $g_z = e^{i k_z/2}$, $z=x,y$. With $\delta \Delta_\mu(\bm{k}) = \partial_{\Delta_\mu^R} \Delta_{\bm{k}}$, the matrix $S^\theta$ then becomes (in the basis $(\overline{\Delta}_{AB,1}, \overline{\Delta}_{AC,1}, \overline{\Delta}_{AB,2}, \overline{\Delta}_{AC,2})^T$)

\begin{align}
    S^\theta = w \begin{pmatrix}
        |f^n_B(\theta, \qpdw)|^2
        & f^n_B(\theta, \qpdw) f^n_C(\theta, \qpdw)^* g_x g_y^* &|f^n_B(\theta, \qpdw)|^2 g_x^2 & f^n_B(\theta, \qpdw) f^n_C(\theta, \qpdw)^* g_x g_y\\
        f^n_B(\theta, \qpdw)^* f^n_C(\theta, \qpdw) g_x^* g_y & |f^n_C(\theta, \qpdw)|^2 &f^n_B(\theta, \qpdw)^* f^n_C(\theta, \qpdw) g_x g_y& f^n_C(\theta, \qpdw)|^2 g_y^2\\
        |f^n_B(\theta, \qpdw)|^2 (g_x^2)^* & f^n_B(\theta, \qpdw) f^n_C(\theta, \qpdw)^* g_x^* g_y^*
        &|f^n_B(\theta, \qpdw)|^2
        &f^n_B(\theta, \qpdw) f^n_C(\theta, \qpdw)^* g_x^* g_y\\
        f^n_B(\theta, \qpdw)^* f^n_C(\theta, \qpdw) g_x^* g_y^*
        &f^n_C(\theta, \qpdw)|^2 (g_y^2)^*
        &f^n_B(\theta, \qpdw)^* f^n_C(\theta, \qpdw) g_x g_y^*
        &|f^n_C(\theta, \qpdw)|^2
    \end{pmatrix},
\end{align}
\end{widetext}
where $w = -J(n_F(ck_0) - \frac{1}{2})/4ck_0 > 0$. Note that $S^\theta$ does not depend on $f^m_B(\theta)$ or $f^m_C(\theta)$, because due to the nearest-neighbor interaction, the $B$ and $C$ orbitals can only pair with $A$, but there is no contribution from $A$ on the flat band ($f^n_A = 0)$. Next, we integrate $S^\theta$ over $\theta$, assuming that $k_0$ is small enough that the $g_z$ terms can be approximated as constants with respect to $\theta$. Diagonalizing the integrated $S^\theta$ yields eigenvalues

\begin{align}
    \varepsilon_1 &= 0\\
    \varepsilon_2 &= 0\\
    \varepsilon_{3,4} &= w \pm w\sqrt{[F^n_{BB}(\bm{q}) - F^n_{CC}(\qpdw)]^2 + 4 |F^n_{BC}(\bm{q})|^2},
\end{align}
where

\begin{align}
    F^n_{\alpha \beta}(\qpdw) = \frac{1}{2\pi} \int_0^{2\pi} f^n_\alpha(\theta, \qpdw)^* f^n_\beta(\theta, \qpdw) \intd{\theta}.
\end{align}
By the Cauchy-Schwarz inequality, we have

\begin{align}
    |F^n_{BC}(\bm{q})|^2 \leq F^n_{BB}(\bm{q}) F^n_{CC}(\bm{q}).
\end{align}
As $w > 0$, the maximal eigenvalue is achieved when this inequality is tight, in which case the eigenvalues become $\varepsilon_3 = 2w, \varepsilon_4 = 0$, as seen by using $F^n_{BB}(\bm{q}) + F^n_{CC}(\bm{q}) = 1$, which follows from the normalization of the Bloch states. This occurs when $f^n_B(\theta, \qpdw)$ and $f^n_C(\theta, \qpdw)$ are linearly dependent. One such case is when $f^n_B(\theta, \qpdw)$ and $f^n_C(\theta, \qpdw)$ are constant with respect to $\theta$. In the worst case, for example when $f^n_B(\theta, \qpdw) = \sin \theta, f^n_C(\theta, \qpdw) = \cos \theta$, we have $|F^n_{BC}(\bm{q})|^2 = 0$ and $F^n_{BB}(\bm{q}) - F^n_{CC}(\qpdw) = 0$, leading to eigenvalues $\varepsilon_3 = \varepsilon_4 = w$.

This result can be used to interpret the $2\qpdw = \bm{M}$ PDW state of the Lieb lattice. For the whole Brillouin zone, the functions $f_B^n, f_C^n$ are given by $f_B^n(\bm{k}) = 2 \cos (k_y/2)/\epsilon_{\bm{k}m}$ and $f_C^n(\bm{k}) = -2 \cos (k_x/2)/\epsilon_{\bm{k}m}$, where the normalization factor $\epsilon_{\bm{k}m} = 2\sqrt{\cos^2(k_x/2) + \cos^2(k_y/2)}$ such that $\pm \epsilon_{\bm{k}m}$ also gives the dispersion of the dispersive bands. When $2\qpdw = \bm{M}$, the main pairing contribution comes from the Dirac point interacting with the center of the flat band. Since $\cos x \approx 1$ to first order for small $x$, this is where $f^n_B$ and $f^n_C$, and thus the orbital composition of the FB states (determined by $|f^n_B|^2$ and $|f^n_C|^2$) are at their most stable as seen in Fig.~\ref{fig:lieb_results}b of the main text.  On the other hand, using $\cos(x - \pi/2) = \sin x$ and $\sin x \approx x$ for small $x$ we obtain $f^n_B(\theta) \approx \sin \theta $ and $f^n_C(\theta) \approx \cos \theta$ on a small circle around $(\pi, \pi)^T$, which would be the main pairing region on the flat band with $\bm{q} = 0$. This is also reflected by the rapidly changing orbital composition on the flat band around the Dirac points in Fig.~\ref{fig:lieb_results}b. Thus, $2\qpdw = \bm{M}$ is the most, and $\qpdw = 0$ the least favored. This result arises from the off-diagonal elements of the pairing susceptibility, and in particular as an interference effect from such off-diagonal derivatives $\partial_{\Delta_\mu} \partial_{\Delta_\nu} E$ of the BdG energies $E$ where $\Delta_\mu$ and $\Delta_\nu$ are order parameters for different orbital pairs (A-B and A-C).

Some caveats to the above calculation should be noted. Here we have only integrated the pairing susceptibility along a thin circle near the Dirac point. In a more accurate calculation, one should also integrate along the radial coordinate $k_0$ (before the diagonalization). In addition, considering only a small region of the Brillouin zone around the Fermi level is not quantitatively correct here, since a significant portion of the integral is contributed by states that are far away from the Fermi level due to the vanishing of the DOS contribution of the Dirac points (and suppressed intra-flat-band pairing). Nevertheless, this result offers a qualitative explanation for why the PDW state is favored.

\subsection{Strong-coupling Limit}

\label{sec: supp_lieb_strong_coupling_limit}

When comparing the Lieb lattice flat band ($\mu = 0$) and vHs ($\mu = 2)$ with nearest-neighbour interactions, we find that the former has significantly weaker dependence on $\qpdw$. This is visible in Figs. \ref{fig:suppl_lieb_results}(c)-(d); changing $\qpdw$ results in fluctuations of only order $10^{-3}$ in the grand potential $\Omega$ for the flat band, but of order $10^{-1}$ for the vHs. This effect also directly leads to smaller superfluid weights and BKT temperatures for the flat band.  The difference can be understood by looking at the strong-coupling limit. For the flat band ($\mu = 0$) and in the limit $t = 0$, where $t$ is the hopping amplitude, the BdG Hamiltonian \eqref{eq_supp: H_bdg} can be diagonalized exactly, yielding a degenerate solution $\Delta = J/2$, where

\begin{align}
    \Delta \equiv \sqrt{|\overline{\Delta}_{AB}|^2 + |\overline{\Delta}_{AC}|^2},
\end{align}
where $\overline{\Delta}_{AB}$ is the order parameter in either the positive or negative $x$-direction towards orbital $B$ from the central orbital $A$,  (see Fig.~\ref{fig:lieb_results}(a) in the main text for the orbital naming convention), with the other one set to zero; and similarly for the orbital pair $A, C$ in the $y$-direction. The solution is degenerate with respect to these choices, as well as to the phases of the order parameters. The corresponding expression for $\Omega$ is

\begin{align}
    \Omega(t=0) = -2\Delta + \frac{2}{J} \Delta^2.
\end{align}
Note that this result has no dependence on $\bm{q}$, since only the diagonal blocks of the BdG Hamiltonian depend on $\bm{q}$, and they are set to zero in the $t = 0$ limit. By a straight-forward but lengthy application of Eq.~\eqref{eq_supp: omega_derivative_general}, we can expand in powers of $t$ and find

\begin{align}
    \Omega = -2\Delta + \frac{2}{J} \Delta^2 - \frac{7}{2\Delta} t^2 + \mathcal{O}(t^4/\Delta),
\end{align}
where the odd terms vanish due to the particle-hole symmetry of bipartite lattices. Notably, contrary to what might be expected in general, the coefficient that is second order in $t$ also doesn't depend on $\bm{q}$. Numerically, we find that the $\qpdw$-dependence only appears in the sixth order correction. Since the PDW state for the flat band only appears for $J > J_c \approx 2.9t$, this explains why its dependence on $\qpdw$ is so weak. In contrast, for the vHs we find numerically that $\qpdw$-dependence appears in second-order, leading to a larger fluctuation of the grand potential, and subsequently larger superfluid weights and BKT temperatures (see Fig.~\ref{fig: bkt_temperatures} in the main text). 

\section{Additional Results for the Kagome Lattice}

\label{sec: supp_kagome_results}

\begin{figure*}
    \centering
    \includegraphics[width=\linewidth]{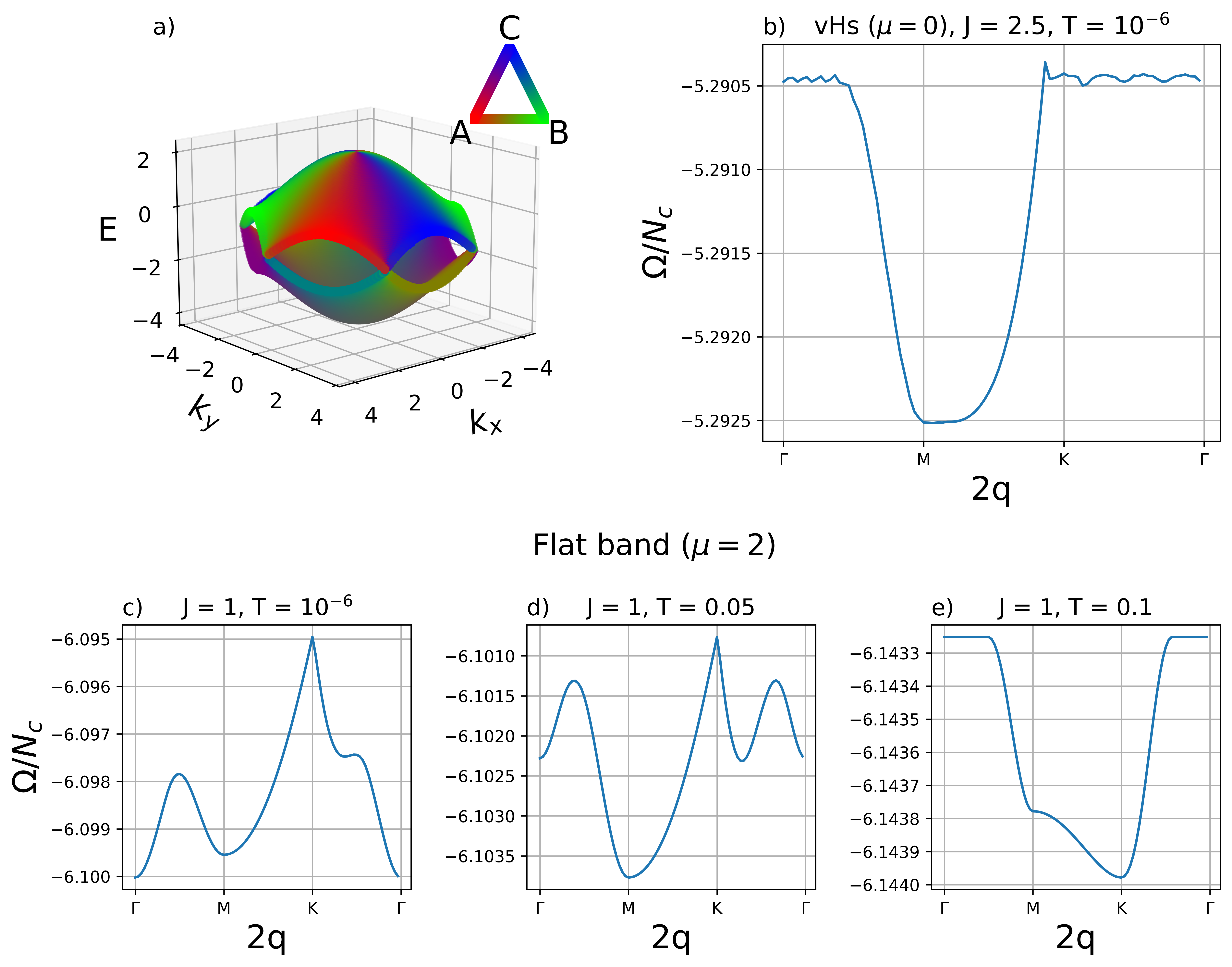}
    \caption{Additional results for the kagome lattice with nearest-neighbour interactions. (\textbf{a}) The band structure of the lattice without the flat band, showing the dispersive bands more clearly (the full band structure is given in Fig.~\ref{fig:kagome_fb_results}b of the main text), with color indicating the orbital composition of each state. On the upper dispersive band, the states are mostly localized to a single orbital throughout the lines from the $\bm{\Gamma}$-point to the $\bm{M}$-points, with full localization at the $\bm{M}$-points. (\textbf{b}) The grand potential for the vHs, confirming the existence of the PDW state. However, the state is unstable, as seen from the flatness of the curve on the path from $\bm{M}$ to $\bm{K}$. This is the same data that was shown in the inset of Fig. \ref{fig: bkt_temperatures} in the main text, but for completeness we repeat it here. The lower panel shows the grand potential for the kagome flat band with $J=1$, showing a transition of the ground state from $2\qpdw = 0$ (\textbf{c}) to $\bm{M}$ (\textbf{d}) to $\bm{K}$ (\textbf{e}) as temperature is increased.}
    \label{fig:suppl_kagome_results}
\end{figure*}

\begin{figure*}
    \centering
    \includegraphics[width=0.7\linewidth]{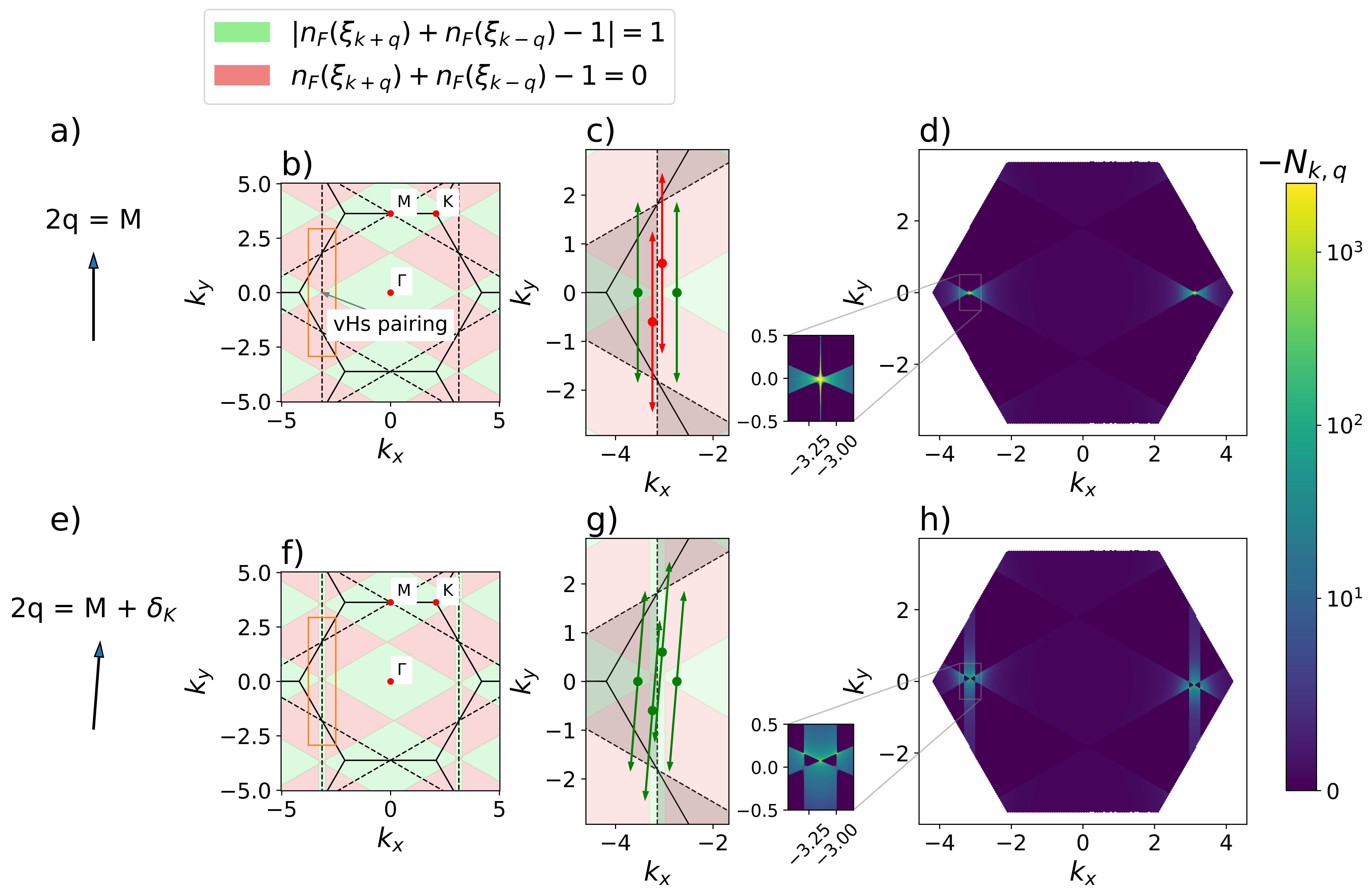}
    \caption{An investigation to the vanishing superfluid weight of the kagome vHs $2 \qpdw = \bm{M}$ (depicted in (\textbf{a})) state by looking at $N_{\bm{k}, \qpdw} \equiv (n_F(\xi_{\bm{k}+\qpdw}) + n_F(\xi_{\bm{k}-\qpdw}) - 1)/(\xi_{\bm{k}+\qpdw} + \xi_{\bm{k}-\qpdw})$, the prefactor of the pairing susceptibility (Eq.~\eqref{eq: pair_susceptibility} in the main text, with the band indices $m, n$ here set to the upper dispersive band that hosts the Fermi surface). When $T \rightarrow 0$ and $n_F$ becomes a step function, the numerator of $N_{\bm{k}, \qpdw}$ vanishes whenever $\bm{k}+\qpdw$ and $\bm{k}-\qpdw$ are on opposite sides of the Fermi surface (FS), shown shaded in red in (\textbf{b}). The solid lines denote Brillouin zone boundaries, while the dashed lines show the FS. In the remaining green regions, most of the contribution to the susceptibility comes from the regions where both $\xi_{\bm{k}+\qpdw}$ and $\xi_{\bm{k}-\qpdw}$ are small, i.e.~near the FS where the orientation of $\qpdw$ and the FS coincide, indicated by the orange rectangle. (\textbf{c}) zooms in on this region, showing how the green region intersects the FS boundary only near $\bm{k} = (\pm \pi, 0)^T$, corresponding to pairing between two vHs points. Here, the darker regions are outside the FS, and the arrows illustrate how $\bm{k}+\qpdw$ and $\bm{k}-\qpdw$ end up on different sides of the FS when moving along the FS boundary. In (\textbf{d}), the full value of $N_{\bm{k}, \qpdw}$ is presented with a small non-zero temperature $T = 10^{-4}$, showing how only the vHs points contribute meaningfully. On the second row, $2 \qpdw$ is changed slightly towards the $\bm{K}$-point by $\delta_{\bm{K}} = (0.15, 0)^T$ (\textbf{e}). This opens a new region of possible pairing along the vertical strip of the FS in (\textbf{f}) and (\textbf{g}). Consequently, $N_{\bm{k}, \qpdw}$ in (\textbf{h}) shows a considerable contribution along this strip. While the peak value of $N_{\bm{k}, \qpdw}$ has decreased, the area where it is considerable has increased. Therefore, the contribution of pairing to the grand potential can remain very similar, that is, the change of the momentum by $\delta_{\bm{K}}$ has a negligible energy cost. This effect may explain
    the instability to order parameter phase fluctuations seen in Fig.~\ref{fig:suppl_kagome_results}b and the vanishing superfluid weight.
    }
    \label{fig:suppl_kagome_vhs_sfw_schematic}
\end{figure*}

\begin{figure*}
    \centering
    \includegraphics[width=0.7\linewidth]{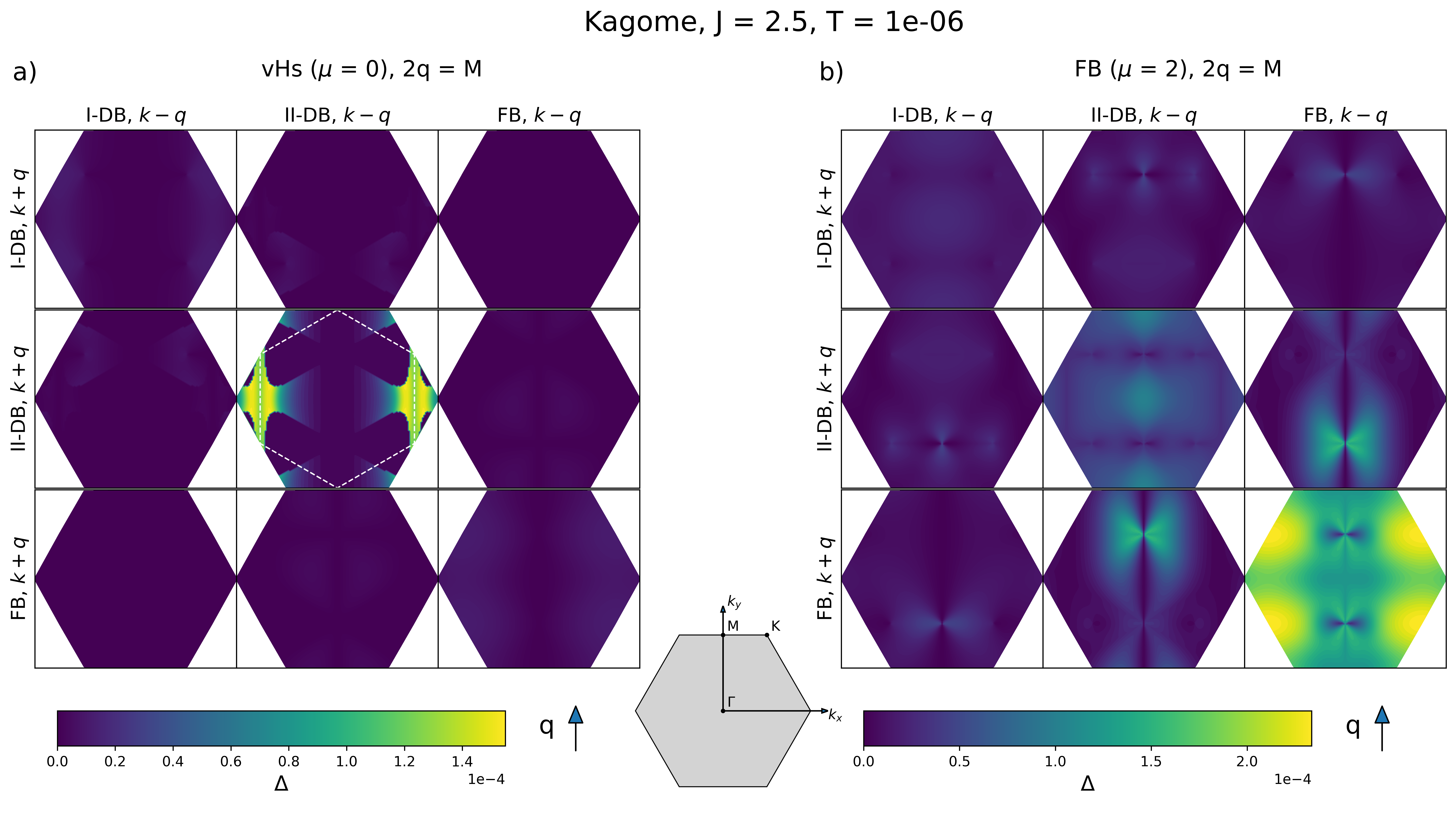}
    \caption{Band and $\bm{k}$-resolved pairing over the Brillouin zone (see Eq.~\eqref{eq_supp: gap_eq_band_resolved}) for the kagome lattice vHs $2\qpdw = \bm{M}$ state, and the flat band $2\qpdw = \bm{M}$ state (\textbf{b}). The blue arrows depict $\qpdw$. The diagonal elements show intra-band pairings, and the off-diagonal the various inter-band pairings. The lower and upper dispersive bands are dubbed I-DB and II-DB, respectively. The color at $\bm{k}$ indicates the strength of pairing between $\bm{k} + \qpdw$ and $\bm{k} - \qpdw$. The white dashed line in (\textbf{a}) indicates the Fermi surface; in (\textbf{b}) there is no well-defined Fermi surface due to the flat band. Contributions from different real-space order parameters $\overline{\Delta}_{0 \alpha j \beta}$ have been combined by taking absolute values and summing over them to obtain an aggregate band and $\bm{k}$-resolved image. In (\textbf{a}), most of the contribution comes from intra-band pairing in the region where $\bm{q}$ lines up with the Fermi surface; see also Fig.~\ref{fig:suppl_kagome_vhs_sfw_schematic}.}
    \label{fig: suppl_kagome_M_delta_band_k_resolved}
\end{figure*}

\begin{figure*}
    \centering
    \includegraphics[width=\linewidth]{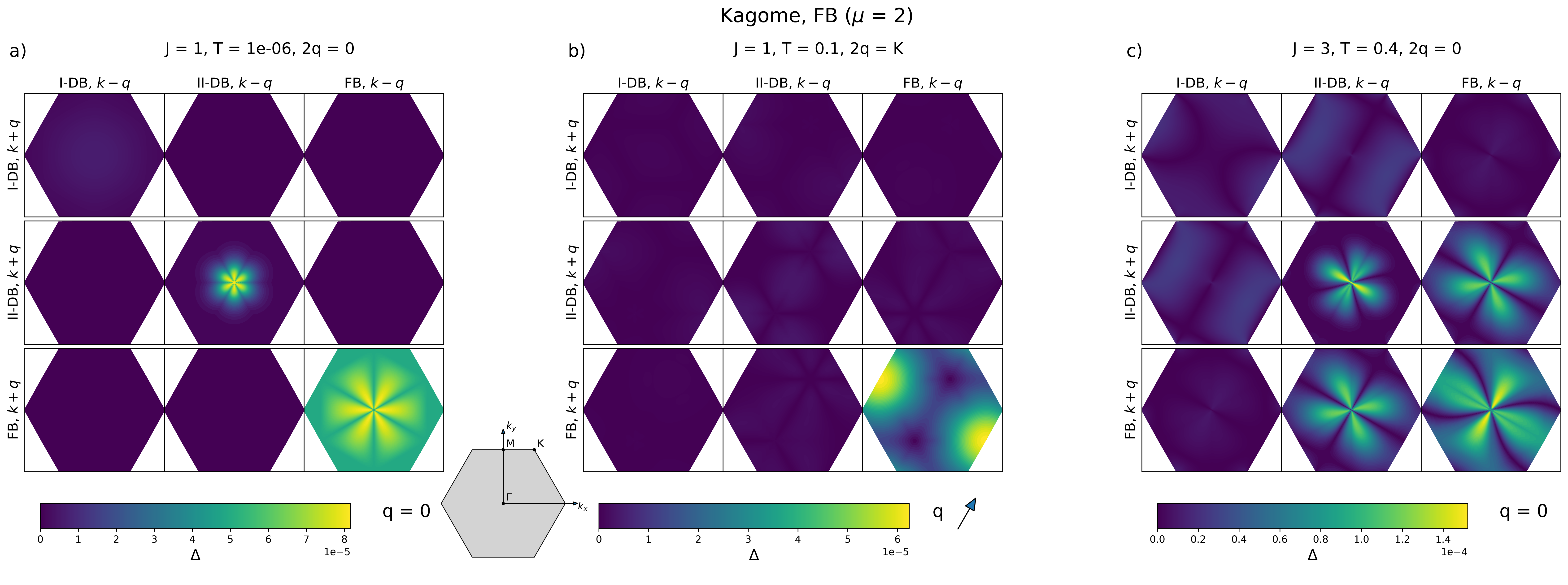}
    \caption{Band and $\bm{k}$-resolved pairing over the Brillouin zone (see Eq.~\eqref{eq_supp: gap_eq_band_resolved}) for different phases of the kagome lattice flat band (see Fig.~\ref{fig:kagome_fb_results} in the main text): the low-temperature, weak-interaction $2\qpdw = 0$ state (\textbf{a}); the high-temperature, weak-interaction $2\qpdw = \bm{K}$ state (\textbf{b}); and the high-temperature, strong-interaction $\qpdw = 0$ state (\textbf{c}). The remaining $2\qpdw = \bm{M}$ phase is shown in Fig.~\ref{fig: suppl_kagome_M_delta_band_k_resolved}. The blue arrow in $(\textbf{b})$ depicts $\qpdw$. The diagonal elements show intra-band pairings, and the off-diagonal the various inter-band pairings. The lower and upper dispersive bands are dubbed I-DB and II-DB, respectively. The color at $\bm{k}$ indicates the strength of pairing between $\bm{k} + \qpdw$ and $\bm{k} - \qpdw$. Contributions from different real-space order parameters $\overline{\Delta}_{0 \alpha j \beta}$ have been combined by taking absolute values and summing over them to obtain an aggregate band and $\bm{k}$-resolved image.}
    \label{fig: suppl_kagome_G_K_delta_band_k_resolved}
\end{figure*}

For the kagome lattice flat band ($\mu = 2$) with nearest-neighbour interactions, we find the rich phase diagram with four different superconducting phases that was shown in Fig.~\ref{fig:kagome_fb_results}(c) of the main text. In Figs.~\ref{fig:suppl_kagome_results}(c)-(e), we show the evolution from $\qpdw = 0$ to $2\qpdw = \bm{M}$ to $2\qpdw = \bm{K}$ in terms of the grand potential $\Omega$ as the temperature is increased. In Fig.~\ref{fig:suppl_kagome_results}(b), we repeat for completeness the graph of $\Omega$ for the vHs that was already shown in the inset of Fig. \ref{fig: bkt_temperatures} of the main text. The flatness of the curve near its minimum indicates that the $2\qpdw = \bm{M}$ ground state is unstable to fluctuations of $\qpdw$, giving an extremely small superfluid weight. We find that this instability makes the system very susceptible to finite-size effects and that the superfluid weights are either zero or at least orders of magnitude smaller than for any other system considered in this work. Fig.~\ref{fig:suppl_kagome_vhs_sfw_schematic} discusses why the superfluid weight of the PDW vanishes at the kagome vHs by considering contributions to pairing close to the Fermi surface. The band and $\bm{k}$-resolved pairing strength for the PDW state is shown in Fig.~\ref{fig: suppl_kagome_M_delta_band_k_resolved}(a), showing strong concentration to the Fermi the surface and agreeing with the insights of Fig.~\ref{fig:suppl_kagome_vhs_sfw_schematic}. The same data for the various flat band ($\mu = 2$) states are shown in Figs.~\ref{fig: suppl_kagome_M_delta_band_k_resolved}(b) and \ref{fig: suppl_kagome_G_K_delta_band_k_resolved}.


Finally, the size of the order parameters and superfluid weights are shown in Fig.~\ref{fig:suppl_deltas_sfws} alongside the Lieb lattice results. An interesting observation is that as found in the main text, for the kagome lattice with on-site interactions and $J \gtrsim 1.8$ (with $J$ understood as the Hubbard-$U$), the BKT temperature is higher for the vHs than for the FB (see Fig.~\ref{fig: bkt_temperatures}), despite the order parameters of the FB being larger (see Fig.~\ref{fig:suppl_deltas_sfws}a). An explanation can be found in Fig.~\ref{fig:suppl_deltas_sfws}b. As the superfluid weight is typically decreasing as a function of temperature, we find that for a large BKT temperature, a system needs both a large superfluid weight at $T=0$ as well as a sizable $T_c$ so that the superfluid doesn't decrease too quickly. In Fig.~\ref{fig:suppl_deltas_sfws}b we see that, for $J=1$, the kagome vHs with on-site interactions has a very large superfluid weight compared to the FB at zero temperature, but the order parameters, and hence $T_c$, are so small that the BKT temperature ends up being small as well. However, when $J$ is increased, the large zero-temperature superfluid weight eventually allows the vHs to overtake the FB when its $T_c$ increases.

\clearpage

\bibliography{rvb_paper3,bibfile_references,bibfile_02}


\end{document}